\documentclass[aps,prb,twocolumn,amsmath,amssymb,floatfix,superscriptaddress]{revtex4-2}
\usepackage{graphicx}% include figure files
\usepackage{bm}% bold math
\usepackage[hidelinks]{hyperref}% add hypertext capabilities
\usepackage{braket,bbold,color,empheq}
\usepackage{tcolorbox}

\newcommand{\bs}[1]{\boldsymbol{#1}}
\newcommand{\normord}[1]{:\mathrel{#1}:}
\begin{document}

\title{Exact Quantum Scars in the Chiral Non-Linear Luttinger Liquid}

\author{Frank Schindler}
\affiliation{Princeton Center for Theoretical Science, Princeton University, Princeton, NJ 08544, USA}

\author{Nicolas Regnault}
\affiliation{Department of Physics, Princeton University, Princeton, NJ 08544, USA}

\author{B. Andrei Bernevig}
\affiliation{Department of Physics, Princeton University, Princeton, NJ 08544, USA}
\affiliation{Donostia International Physics Center, P. Manuel de Lardizabal 4, 20018 Donostia-San Sebastian, Spain}
\affiliation{IKERBASQUE, Basque Foundation for Science, Bilbao, Spain}

\begin{abstract}
While the chiral linear Luttinger liquid is integrable via bosonization, its non-linear counterpart does not admit for an analytic solution. In this work, we find a sub-extensive number of exact eigenstates for a large family of density-density interaction terms. These states are embedded in a continuum of strongly-correlated excited states. The real-space entanglement entropy of some exact states scales logarithmically with system size, while that of others has volume-law scaling. We introduce momentum-space entanglement as an unambiguous differentiator between these exact states and the remaining excited states. With regard to momentum space, the exact states behave as bona fide quantum many body scars: they exhibit identically zero momentum-space entanglement, while typical eigenstates behave thermally. We corroborate this finding by a level statistics analysis. Furthermore, we detail the general formalism for systematically finding all interaction terms and associated exact states, and present a number of infinite exact state sequences extending to arbitrarily high energies. Unlike many previous examples of quantum many body scars, the exact states uncovered here do not lie at equidistant energies, and do not follow from a special operator algebra. Instead, they are uniquely enabled by the interplay of Fermi statistics and chirality.
\end{abstract}

\maketitle

\section{Introduction} 
A core tenet of statistical physics is that any isolated many-body system that is sufficiently complex eventually reaches thermodynamic equilibrium. In quantum physics, the time evolution of Hamiltonian eigenstates is trivial in that it merely consists in a phase factor. This assumption gives rise to the eigenstate thermalization hypothesis (ETH): in the thermodynamic limit, generic eigenstates should become locally indistinguishable from thermal density matrices~\cite{PhysRevA.43.2046,srednicki1994chaos,rigol2008thermalization,polkovnikov2011colloquium,beugeling2014finite,kim2014testing,d2016quantum,garrison2018does}.
However, the notion of what constitutes a ``sufficiently complex" many-body system is subtle. For instance, some integrable classical and quantum many-body systems, which are characterized by an extensive number of conserved quantities, fail to thermalize. Moreover, thermalization may be inhibited even in more realistic and less fine-tuned systems when strong disorder is present, as exemplified by Anderson localization~\cite{Anderson58}. This concept can be generalized to strongly correlated many-body systems, giving rise to many-body localized (MBL) phases~\cite{PhysRevLett.95.206603,BASKO20061126,PhysRevB.77.064426,PhysRevB.82.174411,PhysRevLett.109.017202,PhysRevLett.110.260601,nandkishore2015many,abanin2018review}, whose fate in the thermodynamic limit is still under investigation~\cite{deroeck2017stability, suntajs2020quantum,sels2020dynamical,crowley2020constructive}. These are similar to integrable models in that they host an extensive number of \emph{approximately} conserved quantities, but do not require fine tuning.

Recently, the dichotomy between thermalizing many-body systems satisfying the ETH and those with an extensive number of exact or approximate conserved quantities has been revised by the discovery of quantum many-body scars (QMBS)~\cite{bernien2017probing,turner2017quantum,vafek2017entanglement,turner2018quantum, ho2018periodic,mondaini2018comment,moudgalya2018a,moudgalya2018b,iadecola2018exact,castro2018entanglement,sala2019ergodicity,khemani2019int,schecter2019weak,chattopadhyay2019quantum,Zlatko19,Iadecola19,Seulgi19,Motrunich19,Khemani20,odea2020from,alhambra2020revivals,moudgalya2020eta,mark2020eta,mark2020unified,iadecola2020quantum,Shibata20,Michailidis20,Bull20,Kyungmin20,Klebanov20,Mukherjee20,Zhao20,Hudomal20,Hsieh20,Aidelsburger21,Wildeboer21,Giuliano21,surace2021quantum,Kuno21,Chertkov21,mondal2021quantum,pakrouski2021group,pilatowskycameo2021identification,richter2021anomalous,yao2021quantum,ren2021deformed,desaules2021quantum,tang2021multimagnon,jepsen2021catching,martin2021scar,Hongzheng21,Banerjee21}, see also the reviews Refs.~\onlinecite{SerbynReview,PapicReview,SanjayReview}. These form a sub-extensive set of non-thermal eigenstates of a Hamiltonian without conservation laws, or alternatively, of a Hamiltonian restricted to a Hilbert space subsector that resolves all existing conservation laws. In several cases, they are exact eigenstates, \emph{i.e.}, their expression can be derived analytically, whereas the vast majority of eigenstates -- in the thermodynamic limit -- can only be obtained by the practically impossible diagonalization of a Hamiltonian matrix that is exponentially large in the system size. Exact eigenstate QMBS form an integrable subspace that is embedded in a continuum of eigenstates which otherwise satisfy the ETH.
Previous mechanisms for obtaining QMBS involve spectrum-generating algebras~\cite{vafek2017entanglement,BerislavSGA19,moudgalya2020eta,mark2020unified,ren2020quasisymmetry,odea2020from}, or embedding exact subspaces in thermalizing ensembles to construct bespoke QMBS Hamiltonians~\cite{mori2017eth,shiraishimorireply,schecter2019weak,mark2020eta}. 

Here, we employ a novel method of constructing QMBS that is based on tuning a generically non-integrable system between two distinct integrable limits~\cite{martin2021scar}. Specifically, we study the chiral non-linear Luttinger liquid (CNLLL)~\cite{Haldane81,Imambekov09,Yacoby10,Imambekov12,Karrasch_2015,Mirlin15,Diehl16,Lamacraft17,Meden19,Biao19,martin2021scar}, a model of interacting fermions in one spatial dimension with non-linear, yet uni-directional, dispersion. This system has two celebrated integrable limits: (1) a free fermion limit where the interaction is turned off, and (2) a free boson limit where the dispersion is linearized. Between these two limits, we find a rich structure of exact eigenstates that survive away from full integrability. This structure comes about due to a destructive interference of scattering processes in Fock space. The surviving exact eigenstates are non-interacting, Slater-determinant states that are energetically embedded in a continuum of strongly-correlated states. Interestingly, and in contrast to previous examples of QMBS, the exact states discussed here always have a low \emph{momentum-space} entanglement entropy~\cite{HaqueSchoutens07_1,HaqueSchoutens07_2,LiHaldane08,Thomale10,Mondragon13,Lundgren14,Andrade_2014} in a sea of eigenstates having a momentum-space entanglement entropy close to the Page value~\cite{Page93,Bianchi19,Murthy19}. On the other hand, their \emph{real-space} entanglement entropy scaling can range from sub-volume to volume law. This observation is consistent with the recent finding that QMBS must not necessarily have low real-space entanglement~\cite{langlett2021rainbow}.

In Sec.~\ref{sec: model}, we revisit the CNLLL Hamiltonian and discuss its two integrable limits. Then, in Sec.~\ref{sec: exactstates}, we derive the conditions on the interaction term that enable exact eigenstates. In Sec.~\ref{sec: generalizable_sequences}, we derive examples of infinite exact state sequences that generalize to arbitrarily large Hilbert space sectors. Sec.~\ref{sec: genprops} explains the role of particle-hole symmetry in the exact state solutions, and provides a systematic method of constructing exact states. Entanglement properties are discussed in Sec.~\ref{sec: entanglement}. The appendix contains ancillary derivations, a level statistics analysis, and a comprehensive tally of exact states arising from minimal constraints.

\section{Chiral Non-Linear Luttinger Liquid} \label{sec: model}
We begin by introducing the spinless CNLLL Hamiltonian and exploring its two integrable limits. In the \emph{free fermion} limit, interactions are turned off. In the \emph{free boson limit}, the single-particle dispersion relation is linearized around the Fermi energy, so that the system can be solved via bosonization~\cite{Haldane81}.
\subsection{Hamiltonian} \label{subsec: hamiltonian_first}
Consider the Hamiltonian 
\begin{equation} \label{eq: fullH}
H = H_\mathrm{kin} + H_\mathrm{int},
\end{equation}
where we have defined
\begin{equation} \label{eq: vanillavanillaHamiltonian}
\begin{aligned}
H_\mathrm{kin} &= \sum_{p} \epsilon(p) c^\dagger_p c_p, \\
H_\mathrm{int} &= \frac{1}{2} \int_{-L/2}^{L/2} \mathrm{d}x \int_{-L/2}^{L/2} \mathrm{d}y \, V(x-y) c^\dagger_x c_x c^\dagger_y c_y.
\end{aligned}
\end{equation}
In $H_\mathrm{kin}$, the single-particle dispersion relation $\epsilon(p)$ is assumed to satisfy $\mathrm{sgn}\, \epsilon(p) = \mathrm{sgn}\, p$. We impose this condition only for a window of \emph{physically accessible} momenta around the Fermi momentum $p_{\mathrm{F}}=0$. Here, all momenta $|p| \ll \Lambda$ are called physically accessible for some cutoff scale $\Lambda$, which might for instance be given by a microscopic lattice spacing $d$ via $\Lambda \sim 1/d$. Later on, we will Taylor expand $\epsilon(p) = v p + a p^2$ to second order for pedagogical reasons, so that the above condition translates to $v>0$ and $|a| \Lambda \ll v$.

In Eq.~\eqref{eq: vanillavanillaHamiltonian}, the momentum $p \in 2\pi \mathbb{Z}/L$ is discrete and unbounded~\cite{DelftSchoeller98}, $x \in (-L/2,L/2]$ is a continuous position variable with periodic boundary conditions, and we have used the spinless fermion creation and annihilation operators 
\begin{equation} \label{eq: ccdagdef}
c^\dagger_p = \frac{1}{\sqrt{L}} \int_{-L/2}^{L/2} \mathrm{d}x \, e^{\mathrm{i} p x} c^\dagger_x, \quad c^\dagger_x = \frac{1}{\sqrt{L}} \sum_p e^{-\mathrm{i} p x} c^\dagger_p.
\end{equation}
These satisfy the canonical anti-commutation relations
\begin{equation} \label{eq: CAR}
\begin{aligned}
\{c_{x}, c^\dagger_{y}\} &= \delta(x - y), \hphantom{\delta_{p q}} \{c_{x}, c_{y}\} = \{c^\dagger_{x}, c^\dagger_{y}\} = 0, \\
\{c_{p}, c^\dagger_{q}\} &= \delta_{p q}, \hphantom{\delta(x - y)} \{c_{p}, c_{q}\} = \{c^\dagger_{p}, c^\dagger_{q}\} = 0.
\end{aligned}
\end{equation}
The Hamiltonian $H$ preserves two symmetries: U(1) phase rotations, implying a conserved total particle number 
\begin{equation} \label{eq: totalparticlenumop}
\hat{N} = \sum_p c^\dagger_p c_p, \quad [\hat{N},H] = 0,
\end{equation}
and translational symmetry, implying a conserved total momentum 
\begin{equation}
\hat{P} = \sum_p p c^\dagger_p c_p, \quad [\hat{P},H] = 0.
\end{equation}
By the assumption that $\mathrm{sgn}\, \epsilon(p) = \mathrm{sgn}\, p$, the non-interacting ground state $\ket{\Omega}$ of $H_\mathrm{kin}$ satisfies
\begin{equation} \label{eq: non-int-groundstate-def}
c^\dagger_p \ket{\Omega} = 0 \quad (p \leq 0), \quad
c_p \ket{\Omega} = 0 \quad (p > 0).
\end{equation}
In principle, these relations only hold rigorously for $|p| \ll \Lambda$, however, we will assume that they extend to $|p| \rightarrow \infty$: this simplification does not affect the long-wavelength physics~\cite{Haldane81}.
We also note that $\ket{\Omega}$ is doubly degenerate with $c_0 \ket{\Omega}$. 
The normal-ordered version of an operator $\mathcal{O}$ is defined as
\begin{equation}
\normord{\mathcal{O}} \equiv \mathcal{O} - \braket{\Omega | \mathcal{O} | \Omega},
\end{equation}
which corresponds to moving all $c_{p>0}$ and $c^\dagger_{p \leq 0}$ to the right of all other operators in $\mathcal{O}$ while taking into account the canonical anti-commutation relations in Eq.~\eqref{eq: CAR}. It follows that $\normord{\hat{N}}\ket{\Omega}=0$ and $\normord{\hat{P}}\ket{\Omega}=0$. From now on, we will work in the $\braket{\hat{N}} = 0$ sector of Hilbert space, which is spanned by Slater-determinant basis states of the form
\begin{equation} \label{eq: Slaterdef}
\ket{\bs{n},\bs{\bar{n}}} = \left(\prod_{p > 0} c^{\dagger n_p}_p \right) \left(\prod_{p \leq 0} c^{\bar{n}_p}_p \right) \ket{\Omega},
\end{equation}
where the particle and hole occupation vectors $\bs{n}$ and $\bs{\bar{n}}$ with elements equal to $0$ or $1$ have an equal number of non-zero entries. We may order these states by their total momentum eigenvalue:
\begin{equation} \label{eq: totalmomstates}
\normord{\hat{P}} \ket{\bs{n},\bs{\bar{n}}} = \left(\sum_{p > 0} p n_p + \sum_{p \leq 0} |p| \bar{n}_p\right) \ket{\bs{n},\bs{\bar{n}}},
\end{equation}
from which it follows that the Hilbert space at fixed total particle number $\braket{\hat{N}} = 0$ and fixed total momentum $\braket{\hat{P}} = P_\mathrm{tot}$ has dimension $\mathcal{P}(L P_\mathrm{tot}/2\pi)$, where $\mathcal{P}(x)$ counts the integer partitions of $x$~\cite{Haldane81,WrightJacobiIdentity65}.

The normal-ordered Hamiltonian in the $\braket{\hat{N}} = 0$ sector is obtained from Eqs.~\eqref{eq: fullH} and~\eqref{eq: vanillavanillaHamiltonian} as
\begin{equation} \label{eq: Hint_vanilla}
\begin{aligned}
\normord{H} &= \normord{H_\mathrm{kin}} + \normord{H_\mathrm{int}}, \\
\normord{H_\mathrm{kin}} &= \sum_{p > 0} \epsilon(p) c^\dagger_p c_p - \sum_{p\leq 0} \epsilon(p) c_p c^\dagger_p, \\
\normord{H_\mathrm{int}} &= \sum_{p>0} \left[V(p) \sum_{qk} c^\dagger_{q+p} c_q c^\dagger_{k-p}c_k \right],
\end{aligned}
\end{equation}
where we have defined
\begin{equation}
V(p) = \frac{1}{L} \int_{-L/2}^{L/2} \mathrm{d}x \, e^{\mathrm{i}p x} V(x),
\end{equation}
and assumed that $V(x) = V(-x)$, implying that $V(p) = V(-p)$. In Eq.~\eqref{eq: Hint_vanilla}, normal-ordering $H_\mathrm{int}$ is required to restrict the sum over momentum transfers $p$ to positive $p > 0$ so that $\braket{\Omega|\normord{H_\mathrm{int}}|\Omega} = 0$. The Hamiltonian in Eq.~\eqref{eq: Hint_vanilla} has two integrable limits, which we discuss in the following.

\subsection{Free fermion limit} \label{sec: freefermionlimit}
The free fermion limit is defined by setting all interaction potentials $V(q)$ to zero in Eq.~\eqref{eq: Hint_vanilla}. $\normord{H|_{V(q)=0}}=\normord{H_\mathrm{kin}}$ then preserves not only the total particle number $\hat{N}$, but also the individual particle numbers per momentum $p$ as measured by the particle number operators $\hat{n}_p = c^\dagger_p c_p$. Concurrently, we may define the hole number operators $\hat{\bar{n}}_p = c_p c^\dagger_p = 1 - \hat{n}_p$. The eigenstates of $\normord{H|_{V(q)=0}}$ are the momentum-space Slater-determinant states of Eq.~\eqref{eq: Slaterdef} and satisfy
\begin{equation} \label{eq: hkin_estates_fermionic}
\begin{aligned}
\hat{n}_p \ket{\bs{n},\bs{\bar{n}}} &= n_p \ket{\bs{n},\bs{\bar{n}}} \quad (p > 0),
\\
\hat{\bar{n}}_p \ket{\bs{n},\bs{\bar{n}}} &= \bar{n}_p \ket{\bs{n},\bs{\bar{n}}} \quad (p \leq 0), 
\end{aligned}
\end{equation}
where $n_{p > 0} \in \{0,1\}$ and $\bar{n}_{p \leq 0} \in \{0,1\}$ are particle and hole occupation numbers, respectively. In this basis, the Hamiltonian matrix
\begin{equation} \label{eq: hkin_fermionic_diag}
\begin{aligned}
\normord{H|_{V(q)=0}} \ket{\bs{n},\bs{\bar{n}}} =\left[\sum_{p > 0} \epsilon(p) n_p - \sum_{p \leq 0} \epsilon(p) \bar{n}_p\right] \ket{\bs{n},\bs{\bar{n}}}
\end{aligned}
\end{equation}
is diagonal. The states $\ket{\bs{n},\bs{\bar{n}}}$ therefore represent the full solution of the free fermion limit.

\begin{figure*}[t]
\centering
\includegraphics[width=\textwidth]{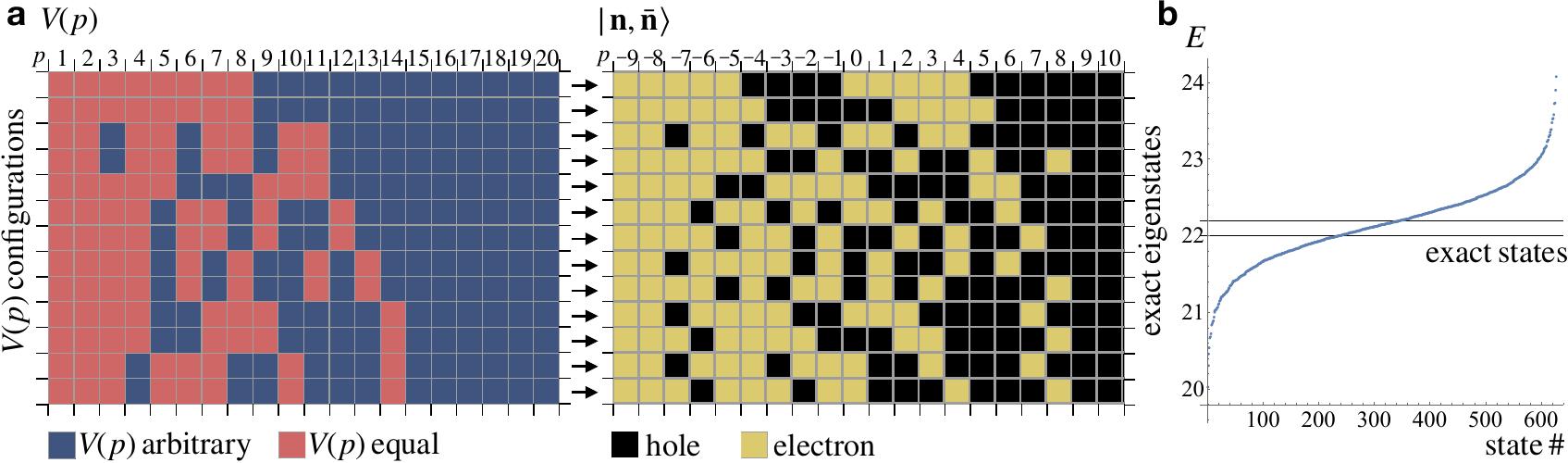}
\caption{Exact eigenstates in the Hilbert space sector with $P_\mathrm{tot} = 20$. (a)~Minimal $V(p)$ tuning patterns (left panel, each row corresponds to a $V(p)$ configuration) giving rise to exact states (right panel, each row corresponds to a different Slater-determinant eigenstate enabled by the $V(p)$ configuration to its left). All $V(p)$'s where $p$ is indicated in red must be set equal for the respective states to become exact eigenstates of the Hamiltonian in Eq.~\eqref{eq: Hint_vanilla}. The exact states are Slater-determinant states where the electronic modes at all momenta indicated in yellow are occupied. All modes at momenta $p<-9$ or $p>10$ are fully occupied or empty, respectively (not shown). For example, setting all $V(p)$ equal for $p \leq 8$ gives rise to two exact states, as shown by the first two rows of the right panel. Tuning less than 8 potentials does not yield any exact states, while tuning more than 8 potentials gives rise to more states than are shown in the right panel (these are further analyzed in Fig.~\ref{fig: scaling}). (b)~Full energy spectrum of $\normord{H}$ in Eq.~\eqref{eq: Hint_vanilla} when restricting to $P_\mathrm{tot} = 20$, sorted by increasing energies. The horizontal axis is the eigenstate index. For this plot, we have set $\epsilon(p) = vp + ap^2$ with $v = 1$ and $a = 0.1/P_\mathrm{tot}$, and have chosen $V(p) = 0.1$ for all momenta highlighted in red in the last row of a), while for all remaining momenta $V(p)$ is sampled uniformly from the range $[0.05,0.15]$. The energies of the two exact states, corresponding to the last two rows in a), are indicated by horizontal lines.}
\label{fig: exactstates}
\end{figure*}

\subsection{Free boson limit} \label{sec: freebosonlimit}
The free boson limit is defined by linearizing the dispersion relation $\epsilon(p) \equiv v p$ in Eq.~\eqref{eq: Hint_vanilla}.
It is then useful to introduce the collective boson operators~\cite{Haldane81,DelftSchoeller98}
\begin{equation} \label{eq: boson_ops_intro}
b_p = \sqrt{\frac{2\pi}{L p}} \sum_k c^\dagger_{k-p} c_k, \quad b^\dagger_p = \sqrt{\frac{2\pi}{L p}} \sum_k c^\dagger_{k+p} c_k,
\end{equation}
which are only defined for positive momenta $p > 0$. These operators obey the commutation relations of boson creation and annihilation operators
\begin{equation}
\begin{aligned}
\relax[b_p, b^\dagger_q] &= \delta_{pq}, \\
[b_p, b_q] &= [b^\dagger_p, b^\dagger_q] = 0, \\
[b_p, \hat{N}] &= [b^\dagger_p, \hat{N}] = 0.
\end{aligned}
\end{equation}
Acting with $b^\dagger_p$, $p>0$, on $\ket{\Omega}$, or on any ground state of $\normord{H_\mathrm{kin}}$ at fixed total particle number, recovers all other states of the many-body Hilbert space that share the same total particle number~\cite{Haldane81}. In particular, the $\braket{\hat{N}} = 0$ sector is spanned by the orthonormal basis
\begin{equation} \label{eq: bosonbasis}
\ket{\bs{m}} = \prod_{p>0} \frac{b^{\dagger m_p}_p}{\sqrt{m_p!}} \ket{\Omega},
\end{equation}
where $0 \leq m_{p} \in \mathbb{Z}$, $p > 0$, are boson occupation numbers. Let us first find the eigenstates of $\normord{H_\mathrm{kin}|_{\epsilon(p) = v p}}$ in the bosonic basis. We use the commutators
\begin{equation} \label{eq: HkinSGA}
\begin{aligned}
\relax \left[\sum_{k} k c^\dagger_k c_k, b_p\right] &= - p b_p, \quad \left[\sum_{k} k c^\dagger_k c_k, b^\dagger_p \right] = p b^\dagger_p,
\end{aligned}
\end{equation}
which have the form of a spectrum-generating algebra (SGA)~\cite{vafek2017entanglement,BerislavSGA19,moudgalya2020eta,mark2020unified,ren2020quasisymmetry,odea2020from}. It then follows that
\begin{equation} \label{eq: hkin_bosonic_diag}
\normord{H_\mathrm{kin}|_{\epsilon(p) = v p}} \ket{\bs{m}} = v\sum_{p>0} p m_p \ket{\bs{m}}.
\end{equation}
This spectrum matches with that of Eq.~\eqref{eq: hkin_fermionic_diag} when restricting to $\epsilon(p) = v p$. Next, one can re-express the normal-ordered interaction term of Eq.~\eqref{eq: Hint_vanilla} as
\begin{equation} \label{eq: boseinteraction}
\normord{H_\mathrm{int}} = \frac{L}{2\pi} \sum_{p>0} V(p) p b^\dagger_p b_p.
\end{equation}
Here, each term is proportional to the boson number operator $b^\dagger_p b_p$, and hence the interaction becomes diagonal in the bosonic representation:
\begin{equation} \label{eq: hint_bosonic_diag}
\normord{H_\mathrm{int}} \ket{\bs{m}} = \frac{L}{2\pi} \sum_{p>0} V(p) p m_p \ket{\bs{m}}.
\end{equation}
Finally, the full Hamiltonian of Eq.~\eqref{eq: Hint_vanilla} is diagonalized as
\begin{equation} \label{eq: fullbosonic_solution}
\normord{H|_{\epsilon(p) = v p}} \ket{\bs{m}} = \sum_{p>0} m_p p \left[v+\frac{L V(p)}{2\pi} \right] \ket{\bs{m}}.
\end{equation}
Its ground state is given by $\ket{\Omega}$ for repulsive $V(p) > 0$. Attractive $V(p) < 0$ with $|V(p)| > 2\pi v/L$ leads to boson condensation and an instability. In this work, we focus on the repulsive case. 
The states $\ket{\bs{m}}$ represent the full solution of the free boson limit. Since $p \in 2\pi \mathbb{Z}/L$, the ground state becomes gapless when $L \rightarrow \infty$: a chiral mode cannot be gapped by interactions. If $V(p)=\bar{V}$, then the interactions renormalize the Fermi velocity. If $V(p)$ is not a constant, Eq.~\eqref{eq: fullbosonic_solution} can be thought of as a new gapless dispersion relation $\tilde{\epsilon}(p) = p \left[v+L V(p)/2\pi \right]$.

\section{Exact eigenstate constraints} \label{sec: exactstates}
Away from the two integrable limits discussed in Secs.~\ref{sec: freefermionlimit} and~\ref{sec: freebosonlimit}, we find a remarkably rich structure of exact eigenstates of the CNLLL Hamiltonian in Eq.~\eqref{eq: Hint_vanilla}. These are non-interacting, Slater-determinant states of the form of Eq.~\eqref{eq: Slaterdef}. The existence of these exact eigenstates only assumes that a subset of all $V(p)$, $p > 0$, is set equal. Since the transformation between the fermionic and bosonic basis is unitary, we can analyze in both representations how the exact Slater-determinant eigenstates of Eq.~\eqref{eq: Hint_vanilla} come about.

For convenience, we will set $L = 2\pi$ in the remainder of the main text and in the appendix, rendering all momenta dimensionless. To obtain physical values, all energies and momenta must then be multiplied by $2\pi/L$.

\begin{figure*}[t]
\centering
\includegraphics[width=\textwidth]{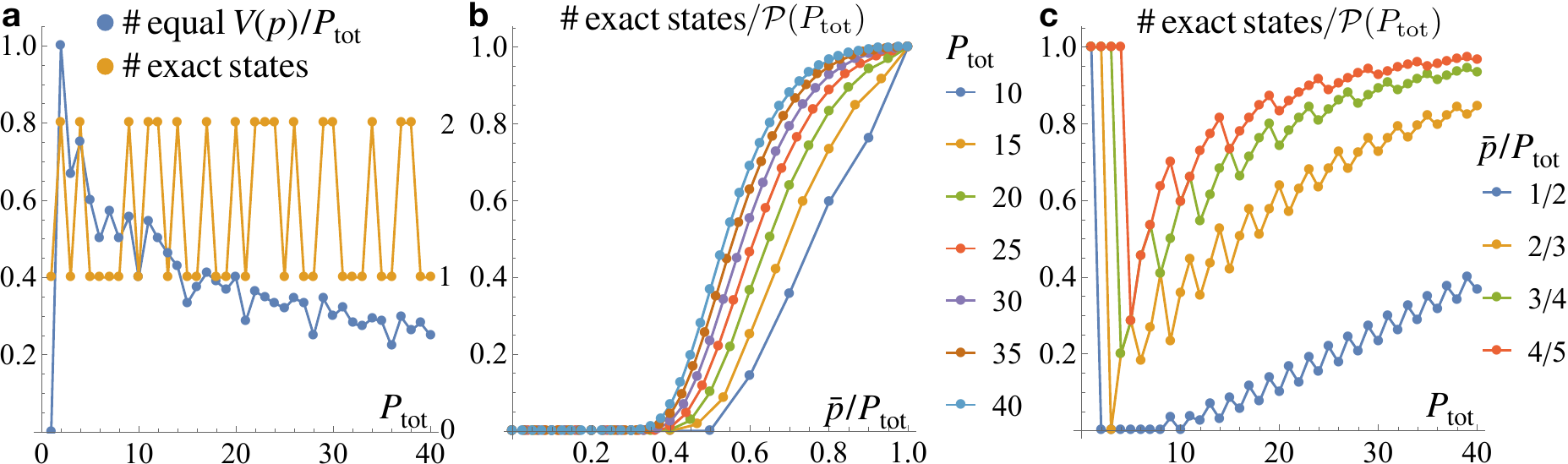}
\caption{Scaling behavior. (a)~Minimal fraction of potentials that need to be set equal to obtain any exact states (blue, left vertical axis) versus resulting number of frugal exact states, which is either $1$ or $2$ (orange, right vertical axis). (b)~Abundance of exact states for homogeneous potential tuning patterns where all $V(p)$ with $p \leq \bar{p}$ are set equal. We provide the ratio of the number of exact states over the Hilbert space dimension $\mathcal{P}(P_\mathrm{tot})$ as a function of the fraction $\bar{p}/P_\mathrm{tot}$ for different total momenta $P_\mathrm{tot}$, which range from 10 to 40. (c)~Fraction of exact states resulting from the tuning pattern where all $V(p)$ with $p \leq \bar{p}$ are set equal, where $\bar{p} = \lceil P_\mathrm{tot}/2 \rceil$, $\bar{p} = \lceil2P_\mathrm{tot}/3 \rceil$, $\bar{p} = \lceil3P_\mathrm{tot}/4 \rceil$, or $\bar{p} = \lceil4P_\mathrm{tot}/5 \rceil$, as a function of $P_\mathrm{tot}$. The asymptotic approach to unity with increasing total momentum, present for all choices of $\bar{p}/P_\mathrm{tot}$, implies asymptotic freedom, \emph{i.e.}, non-interacting physics at high energies.}
\label{fig: scaling}
\end{figure*}

\subsection{Fermionic basis} \label{sec: exactfermi}
The interaction term in Eq.~\eqref{eq: Hint_vanilla} can be rewritten as
\begin{equation} \label{eq: Hint_destructive_interference}
\begin{aligned}
\normord{H_\mathrm{int}}=&\sum_{q>k} \sum_{p > (k-q)/2} \\& \left[V(q-k+p) - \delta_{p \neq 0} V(|p|)\right] c^\dagger_{q+p} c_q c_k c^\dagger_{k-p},
\end{aligned}
\end{equation}
where $q$, $k$, and $p$ are all momenta quantized to integers. Here, $\delta_{p \neq 0}$ is zero when $p=0$, and otherwise unity. For the derivation of Eq.~\eqref{eq: Hint_destructive_interference}, see App.~\ref{app: Hint_derivation}. The contribution from $p=0$ is a diagonal operator in the Fermi basis. Moreover, for all fermionic basis states at fixed total momentum $P_{\mathrm{tot}}$, the single-particle momenta smaller than $-P_{\mathrm{tot}}+1$ and larger than $P_{\mathrm{tot}}$ are fully occupied and empty, respectively, as follows from Eq.~\eqref{eq: totalmomstates}. Hence, all fermionic basis states with total momentum $P_{\mathrm{tot}}$ are annihilated by terms in Eq.~\eqref{eq: Hint_destructive_interference} having $q>P_{\mathrm{tot}}$, $k\leq-P_{\mathrm{tot}}$, or $p > \mathrm{min}(P_{\mathrm{tot}}-q,P_{\mathrm{tot}}-1+k)$. Therefore, $\normord{H_\mathrm{int}}$ has an exact Fermi eigenstate $\ket{\bs{n},\bs{\bar{n}}}$ when, for any triple $q,k,p$ satisfying 
\begin{empheq}{alignat=1}
    &P_{\mathrm{tot}} \geq q>k > -P_{\mathrm{tot}}, \label{eq: qkrange} \\
    &\mathrm{min}(P_{\mathrm{tot}}-q,P_{\mathrm{tot}}-1+k) \geq p > (k-q)/2 \label{eq: prange}
\end{empheq}
($p \neq 0$), either of the following two conditions hold:
\begin{empheq}[left={\empheqlbrace}]{alignat=1}
    c^\dagger_{q+p} c_q c_k c^\dagger_{k-p} \ket{\bs{n},\bs{\bar{n}}} &= 0 \label{eq: condition_on_states} \\
    V(q-k+p) &= V(|p|). \label{eq: condition_on_potentials}
\end{empheq}
Setting all potentials equal, so that $V(p) \equiv \bar{V}$ is constant for all $0 < p \leq P_\mathrm{tot}$, renders all states $\ket{\bs{n},\bs{\bar{n}}}$ of the Hilbert space sector with total momentum $P_\mathrm{tot}$ exact, because in this limit $\normord{H_\mathrm{int}}$ becomes a diagonal matrix in this sector -- equivalently, Eq.~\eqref{eq: condition_on_potentials} is always satisfied for $q,k,p$ in the ranges of Eqs.~\eqref{eq: qkrange} and~\eqref{eq: prange}. Moreover, for the states $\ket{\Omega}$ and $c^\dagger_{1} \ket{\Omega}$, Eq.~\eqref{eq: condition_on_states} is satisfied for \emph{all} choices of $q,k,p$, so that these states remain exact eigenstates for any choice of interaction potential. This can be understood by noting that the total momentum of these states is $P_\mathrm{tot} = 0$ and $P_\mathrm{tot} = 1$, respectively, which gives a one-dimensional Hilbert space sector in both cases -- the number of integer partitions is $\mathcal{P}(0) = \mathcal{P}(1) = 1$, see the discussion below Eq.~\eqref{eq: totalmomstates} and recall that we have set $L=2\pi$. 

More interestingly, we find that there exist states in higher $P_\mathrm{tot}$ sectors where Eq.~\eqref{eq: condition_on_states} holds for all $q$ and $k$ in the range of Eq.~\eqref{eq: qkrange}, but only for a subset of $p$'s in the range of Eq.~\eqref{eq: prange}. This property is unique to chiral fermion states, which have all momenta that are sufficiently small ($p \leq -P_\mathrm{tot}$) or large ($p > P_\mathrm{tot}$) fully occupied and fully empty, respectively, so that the Hamiltonian in Eq.~\eqref{eq: Hint_destructive_interference} restricts to the ranges in Eqs.~\eqref{eq: qkrange} and~\eqref{eq: prange}. As a minimal example, consider the fermionic basis states spanning the Hilbert space sector with $P_\mathrm{tot} = 3$:
\begin{equation}
\ket{\phi_1}=c^\dagger_3 c_0 \ket{\Omega}, \quad \ket{\phi_2}=c^\dagger_2 c_{-1} \ket{\Omega}, \quad \ket{\phi_3}=c^\dagger_1 c_{-2} \ket{\Omega}.
\end{equation}

\begin{widetext} \noindent
In this basis, the interaction term of Eq.~\eqref{eq: Hint_destructive_interference} becomes the $3 \times 3$ matrix
\begin{equation} \label{eq: explicitp3matrix}
\braket{\phi_i | \normord{H_\mathrm{int}} | \phi_j} = 
\begin{pmatrix}
V(1)+V(2)+V(3) & V(1) - V(3) & V(3) - V(2) \\
V(1) - V(3) & 2 V(1) + V(3) & V(1) - V(3) \\
V(3) - V(2) & V(1) - V(3) & V(1) + V(2) + V(3) 
\end{pmatrix}_{ij},
\end{equation}
\end{widetext}
where we have used Eqs.~\eqref{eq: CAR} and~\eqref{eq: non-int-groundstate-def}.
We see that $\ket{\phi_2}$ is special, in that the scattering matrix elements in Eq.~\eqref{eq: explicitp3matrix} that act on this state only involve $V(1)$ and $V(3)$, but not $V(2)$. This is because, for this state, Eq.~\eqref{eq: condition_on_states} is satisfied for $p=2$ and \emph{irrespective of} $q$, $k$ in the whole range of Eq.~\eqref{eq: qkrange}.
On the other hand, Eq.~\eqref{eq: condition_on_states} is not in general satisfied for $p=1$, because the scattering term with $q=2$, $k=0$, and $p=1$ in Eq.~\eqref{eq: Hint_destructive_interference} fails to annihilate $\ket{\phi_2}$.
Therefore, to retain $\ket{\phi_2}$ as an exact eigenstate, we must enforce Eq.~\eqref{eq: condition_on_potentials} for $q=2$, $k=0$, and $p=1$, which amounts to
\begin{equation} \label{eq: P3condition}
V(3) = V(1).
\end{equation}
Indeed, this choice nullifies all off-diagonal matrix elements that act on $\ket{\phi_2}$ in Eq.~\eqref{eq: explicitp3matrix}, rendering it an exact Slater-determinant eigenstate. Concomitantly, $V(2)$ can be chosen freely and induces mixing between $\ket{\phi_1}$ and $\ket{\phi_3}$, giving rise to two correlated eigenstates that are not of Slater-determinant type. This minimal example elucidates how non-interacting and correlated states can coexist away from the free fermion/boson limits.

More generally, we define \emph{frugal} exact states as fermionic basis states satisfying Eqs.~\eqref{eq: condition_on_states} or~\eqref{eq: condition_on_potentials} for all triples $q,k,p$ in the ranges of Eqs.~\eqref{eq: qkrange} and~\eqref{eq: prange} and requiring a minimal number of equal potentials $V(p)$ (we do not know of an analytical expression for this minimal number). In App.~\ref{app: exact_tally}, we solve this constraint numerically and list all frugal exact states in all Hilbert space sectors up to $P_\mathrm{tot} = 40$, which has dimension $\mathcal{P}(40) = 37338$. Importantly, a single consistent choice of $V(p)$ can nucleate exact states across a range of total momentum sectors: fixing $V(p)$ to obtain frugal states at a given $P_\mathrm{tot}$ also induces further, generically non-frugal exact states in Hilbert space sectors at total momentum $P<P_\mathrm{tot}$. Their abundance per Hilbert space sector is further analyzed in App.~\ref{app: exact_tally}.

As a concrete example, for $P_\mathrm{tot} = 20$ with Hilbert space dimension $\mathcal{P}(20) = 627$, Fig.~\ref{fig: exactstates}a shows the 13 possible frugal exact states (corresponding to different rows) that arise when $8$ potentials $V(p)$ are set equal. Tuning less potentials does not yield any exact eigenstates. Fig.~\ref{fig: exactstates}b shows that these states generically lie in the bulk of the energy spectrum within a continuum of correlated states. For this panel, we have chosen an interaction where the potentials highlighted in the last row of Fig.~\ref{fig: exactstates}a (left panel) are set equal, while the others are randomly sampled from a uniform distribution. The resulting energy level statistics~\cite{dyson1962brownian,poilblanc1993poisson,oganesyan2007localization,rigol2008thermalization,Atas13,nandkishore2015many} is of Wigner-Dyson type and has a mean adjacent gap ratio $\bar{r} \approx 0.531$, indicating that most eigenstates indeed thermalize~\cite{Atas13}. For a more detailed level statistics analysis of the CNLLL Hamiltonian, see App.~\ref{app: levelstatistics}.

In Fig.~\ref{fig: scaling}a, we study how many potentials $V(p)$ need to be set equal to obtain any exact states (\emph{i.e.}, the frugal states) as a function of $P_\mathrm{tot}$. More exact states arise when tuning a non-minimal number of potentials to be equal: Fig.~\ref{fig: scaling}b shows how many fermionic basis states are retained as exact eigenstates when all potentials $V(p)$ with $p \leq \bar{p}$ are set equal, as a function of $\bar{p}$ and for different values of $P_{\mathrm{tot}}$. Finally, Fig.~\ref{fig: scaling}c shows how many exact states arise in each $P_\mathrm{tot}$ sector when a fixed fraction of potentials with respect to $P_{\mathrm{tot}}$ is set equal (see caption of Fig.~\ref{fig: scaling}). From this figure, we can conclude that as $P_\mathrm{tot}$, and therefore the Hilbert space size, is increased, a smaller fraction of potentials must be fine-tuned to obtain the same fraction of exact states. When the fraction of fine-tuned potentials is held fixed as $P_\mathrm{tot}$ increases, the system exhibits \emph{asymptotic freedom}: in the high-energy limit of large $P_\mathrm{tot}$, almost all eigenstates become non-interacting.

The rich structure of potential tuning patterns and exact states presented here is inherent to chiral systems. As soon as mixing with an anti-chiral mode is allowed, it becomes possible to scatter the highest-momentum electron in the right-moving branch (Fermi velocity $v_{\mathrm{R}} > 0$) with the lowest-momentum electron in the left-moving branch (Fermi velocity $v_{\mathrm{L}} < 0$) by transferring an arbitrary momentum $p$ ($-p$), $p > 0$, to the right-moving (left-moving) electron. Therefore, all eigenstates become at least weakly correlated as soon as chirality is forfeited.

\subsection{Bosonic basis} \label{sec: exact_boson}
The bosonic basis provides an alternative point of view on the physical origin of the exact eigenstates derived in Sec.~\ref{sec: exactfermi}. For this, we rewrite the Hamiltonian of Eq.~\eqref{eq: Hint_vanilla} as 
\begin{equation} \label{eq: bosonic_ham_rewrite}
\begin{aligned}
\normord{H} =& \sum_{p>0} p \left[v+\frac{L V(p)}{2\pi} \right] b^\dagger_p b_p \\&+ \sum_{p > 0} \tilde{\epsilon}(p) c^\dagger_p c_p - \sum_{p\leq 0} \tilde{\epsilon}(p) c_p c^\dagger_p,
\end{aligned}
\end{equation}
where $\tilde{\epsilon}(p) = \epsilon(p) - v p$ only contains the non-linear contributions to the dispersion relation. These non-linearities act as scattering terms between the bosonic eigenstates $\ket{\bs{m}}$ of Eq.~\eqref{eq: bosonbasis}, spoiling the integrability of the free boson limit.

The bosonic modes $\ket{\bs{m}}$ are collective excitations built from many fermionic basis states $\ket{\bs{n},\bs{\bar{n}}}$ in Eq.~\eqref{eq: Slaterdef}:
\begin{equation} \label{eq: unitarybosefermitrafo_intro}
\ket{\bs{m}} = \sum_{\bs{n},\bs{\bar{n}}} U_{\bs{m};\bs{n},\bs{\bar{n}}} \ket{\bs{n},\bs{\bar{n}}}, \quad \ket{\bs{n},\bs{\bar{n}}} = \sum_{\bs{m}} U^\dagger_{\bs{n},\bs{\bar{n}};\bs{m}} \ket{\bs{m}},
\end{equation}
where the unitary bosonization transformation $U$ diagonalizes $\normord{H}$ in the fermionic basis of Eq.~\eqref{eq: Hint_vanilla} when no non-linearities are present, \emph{i.e.}, $\tilde{\epsilon}(p)=0$. Explicitly,
\begin{equation}
U_{\bs{m};\bs{n},\bs{\bar{n}}} = 
\bra{\Omega} \left(\prod_{p \leq 0} c^{\dagger \bar{n}_p}_p \right) \left(\prod_{p > 0} c^{n_p}_p \right) \left(\prod_{p>0} \frac{b^{\dagger m_p}_p}{\sqrt{m_p!}} \right) \ket{\Omega},
\end{equation}
which can be evaluated by using Eqs.~\eqref{eq: boson_ops_intro} and~\eqref{eq: CAR}.
The transformation $U$ does not mix between Hilbert space sectors with different total momenta, and so is a finite-dimensional unitary matrix in each sector. Crucially, some fermionic basis states $\ket{\bs{n},\bs{\bar{n}}}$ have \emph{exactly zero} overlap with a subset $M_{\ket{\bs{n},\bs{\bar{n}}}}$ of bosonic basis states. That is, for these states
\begin{equation} \label{eq: boseunitaryzero}
U_{\bs{m};\bs{n},\bs{\bar{n}}} = 0 \quad \forall \ket{\bs{m}} \in M_{\ket{\bs{n},\bs{\bar{n}}}},
\end{equation}
Now, to ensure that $\ket{\bs{n},\bs{\bar{n}}}$ remains an eigenstate of the Hamiltonian in Eq.~\eqref{eq: bosonic_ham_rewrite} at non-zero $V(p)$, we must enforce the condition
\begin{equation} \label{eq: boson_condition_on_potentials}
\sum_{p>0} V(p) p m_p = \bar{V} \quad \forall \ket{\bs{m}} \notin M_{\ket{\bs{n},\bs{\bar{n}}}}.
\end{equation}
This is because Eqs.~\eqref{eq: boseunitaryzero} and~\eqref{eq: boson_condition_on_potentials} imply that $\ket{\bs{n},\bs{\bar{n}}}$ is an eigenstate of the interaction term $\normord{H_\mathrm{int}}$ in Eq.~\eqref{eq: boseinteraction}, and therefore also of the full Hamiltonian in Eq.~\eqref{eq: bosonic_ham_rewrite}:
\begin{equation}
\begin{aligned}
\normord{H_\mathrm{int}} \ket{\bs{n},\bs{\bar{n}}} =& \sum_{\bs{m}}\sum_{p>0} V(p) p m_p \ket{\bs{m}}\braket{\bs{m}|\bs{n}, \bs{\bar{n}}} \\ =& \bar{V} \sum_{\bs{m}} U^\dagger_{\bs{n},\bs{\bar{n}};\bs{m}} \ket{\bs{m}} = \bar{V} \ket{\bs{n},\bs{\bar{n}}}.
\end{aligned}
\end{equation}
(Recall that we had set $L/2\pi=1$.)
Eq.~\eqref{eq: boson_condition_on_potentials} is a set of linear constraints on the potentials $V(p)$. Together with Eq.~\eqref{eq: boseunitaryzero}, it represents the exact eigenstate constraints of the CNLLL in the bosonic basis.
As a minimal example, for the Hilbert space sector at $P_{\mathrm{tot}} = 3$ that was also discussed in Sec.~\ref{sec: exactfermi}, the first equation of Eq.~\eqref{eq: unitarybosefermitrafo_intro} becomes
\begin{equation} \label{eq: bosonfermiunitaryexample}
\begin{pmatrix} b^\dagger_{3}\ket{\Omega} \\ b^\dagger_{2} b^\dagger_{1}\ket{\Omega} \\ b^{\dagger 3}_{1}\ket{\Omega} \end{pmatrix} = \frac{1}{\sqrt{6}} 
\begin{pmatrix} 
\sqrt{2} & \sqrt{2} & \sqrt{2} \\
-\sqrt{3} & 0 & \sqrt{3} \\
1 & -2 & 1
\end{pmatrix}
\begin{pmatrix} c^\dagger_{1} c_{-2}\ket{\Omega} \\ c^\dagger_{2} c_{-1}\ket{\Omega} \\ c^\dagger_{3} c_{0}\ket{\Omega} \end{pmatrix}.
\end{equation}
From this, we deduce that $c^\dagger_{2} c_{-1} \ket{\Omega}$ has \emph{exactly zero} overlap with $b^\dagger_{2} b^\dagger_{1} \ket{\Omega}$, so that Eq.~\eqref{eq: boseunitaryzero} is satisfied for $$M_{c^\dagger_{2} c_{-1} \ket{\Omega}} = \{b^\dagger_{2} b^\dagger_{1} \ket{\Omega}\}.$$ To ensure that $c^\dagger_{2} c_{-1} \ket{\Omega}$ remains an exact eigenstate, we must also enforce Eq.~\eqref{eq: boson_condition_on_potentials}. For $\ket{\bs{m}}=b^\dagger_{3}\ket{\Omega}$, we have $m_3=1$ as the only non-vanishing entry of $\bs{m}$, so that $3 V(3) = \bar{V}$. For $\ket{\bs{m}}=b^{\dagger 3}_{1}\ket{\Omega}$, we have $m_1=3$ as the only non-vanishing entry of $\bs{m}$, so that $3 V(1) = \bar{V}$. Therefore,
\begin{equation}
V(3) = V(1),
\end{equation}
which is indeed the same condition as we had found in Eq.~\eqref{eq: P3condition}.
We see that the presence of exact states away from the free fermion and free boson limits is a direct consequence of the fact that bosonization, viewed as a unitary transform acting on the fermionic Hilbert space, does not relate all fermionic states to all bosonic states and vice versa \emph{for a chiral system}. This should be contrasted with, for instance, phononic collective modes in crystalline lattices, which have non-zero overlap with \emph{all} local harmonic oscillations of the lattice.

One might ask if, instead of the fermionic basis states $\ket{\bs{n},\bs{\bar{n}}}$, it is also possible to retain some \emph{bosonic} basis states $\ket{\bs{m}}$ as exact eigenstates of $\normord{H}$ away from the free boson limit. This is not the case: as shown in App.~\ref{app: non-linear-boson}, tuning the \emph{continuous} non-linearity parameter $a$ in $\epsilon(p) = vp + ap^2$ away from zero, which breaks the free boson limit, induces a mixing of all bosonic states in any given Hilbert space sector. Hence, no bosonic basis states (permanents) remain exact eigenstates. (See however Ref.~\onlinecite{martin2021scar}, where it is shown that bosonic states may persist as approximate eigenstates.) On the other hand, the interaction potentials $V(p)$, which break the free fermion limit, form a \emph{discrete} set of continuous parameters and can be tuned independently from one another, giving rise to a discrete set of exact fermionic states.

\section{Simple exact state sequences} \label{sec: generalizable_sequences}
Our analysis in Sec.~\ref{sec: exactstates} was predominantly based on numerically solving the constraints Eqs.~\eqref{eq: condition_on_states} and~\eqref{eq: condition_on_potentials}. Due to the asymptotically exponential scaling of the Hilbert space dimension~\cite{Ramanujan1918},
\begin{equation}
\mathcal{P}(P_{\mathrm{tot}}) \sim \frac{1}{4 \sqrt{3} P_{\mathrm{tot}}} e^{\pi \sqrt{\frac{2 P_{\mathrm{tot}}}{3}}} \text{ as } P_{\mathrm{tot}} \rightarrow \infty,
\end{equation}
where $\mathcal{P}(x)$ is the number of partitions of the integer $x$, this approach becomes cumbersome at large total momenta $P_{\mathrm{tot}}$. It is therefore desirable to develop an analytic understanding of some simple exact state solutions that extend to arbitrary total momenta. In this section, we derive three frugal exact state sequences, as well as an infinite family of non-frugal sequences. Then, in Sec.~\ref{sec: genprops}, we describe a more abstract but general method of constructing exact state solutions that includes all sequences discussed here as special cases.

\begin{figure}[t]
\centering
\includegraphics[width=0.48\textwidth]{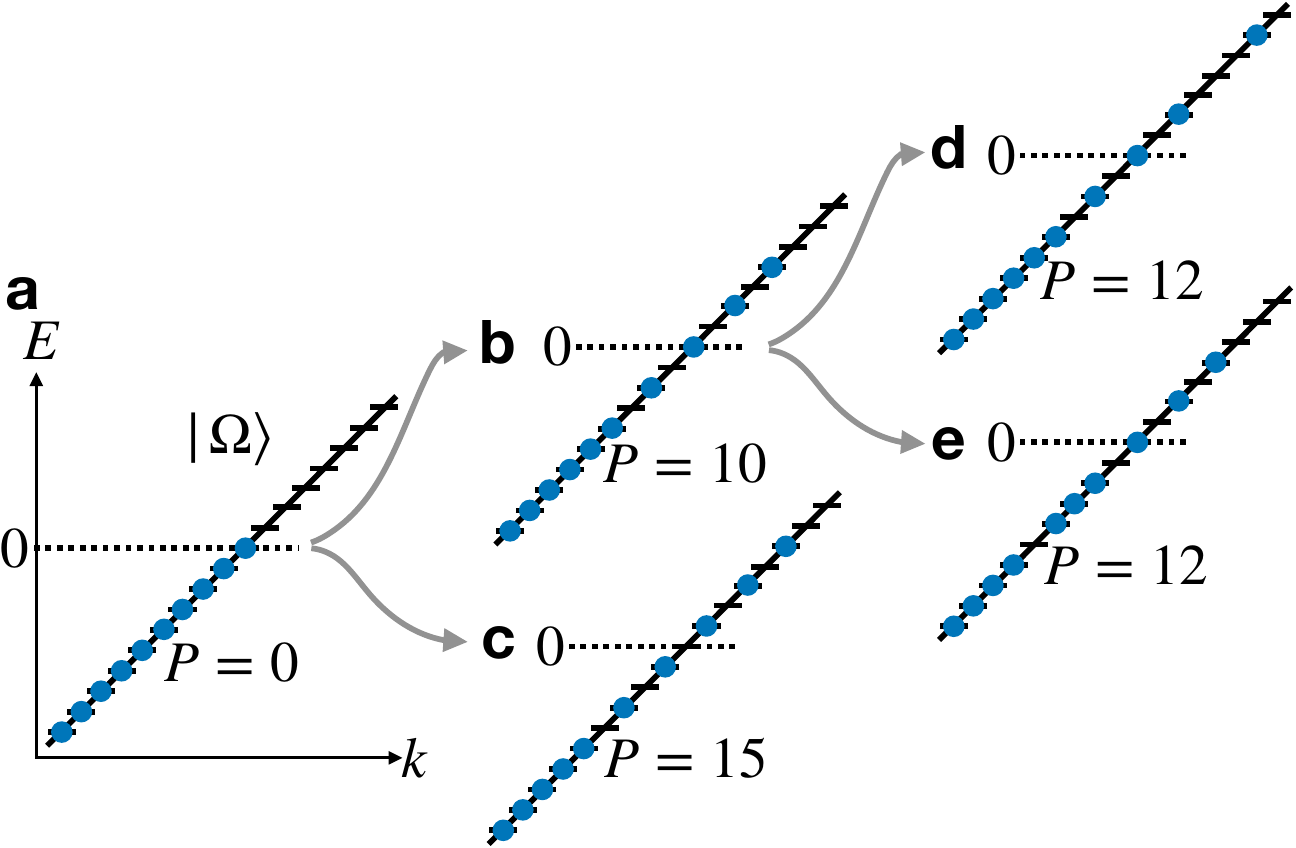}
\caption{Exact state sequences arising from setting the potentials $V(1) = V(3) = \dots = V(2n+1)$ equal. (a)~We start with the filled Fermi sea, given by $\ket{\Omega}$ in Eq.~\eqref{eq: non-int-groundstate-def}. All higher and lower-lying momenta are assumed fully empty and fully occupied, respectively. (b)~For odd $n=2\bar{k}-1$, we move occupied electrons at successive odd momenta $q<0$ to the previously unoccupied momenta at $-q+1$. In the case depicted here, where two electrons are moved (\emph{i.e.}, $\bar{k}=2$), this results in a frugal exact state at total momentum $P_{\mathrm{tot}} = 10$. (c)~For even $n=2(\tilde{k}-1)$, we move occupied electrons at successive even momenta $q<0$ to the previously unoccupied momenta at $-q+1$. In the case depicted here, where three electrons are moved (\emph{i.e.}, $\tilde{k}=3$), this results in a frugal exact state at total momentum $P_{\mathrm{tot}} = 15$. (d)~When additionally $V(2)$ and $V(2(n+1)+1)$ are set equal to $V(1)$, we obtain a descendant state by shifting the electron at the highest occupied momentum of b) by an amount $\Delta=2$. (e)~Another descendant state is obtained by shifting the hole at the lowest unoccupied momentum of b) by $\Delta=-2$. As long as $V(2) = V(1) = V(3) = V(5) = \dots = V(2(n+1)+1)$, both states in d) and e) are exact eigenstates. The same construction can be applied to obtain descendants of c).}
\label{fig: exactfrugals}
\end{figure}

As a first example of an analytical solution, we note that the maximal possible momentum $\bar{p}$ for which we must constrain $V(p)$ to turn a generic fermionic basis state into an exact eigenstate of $\normord{H}$ in Eq.~\eqref{eq: Hint_vanilla} is set by the occupied electron with the largest momentum $p_\mathrm{max}$, and the occupied hole with the lowest momentum $p_\mathrm{min}$, via the relationship $\bar{p} = p_\mathrm{max} - p_\mathrm{min}$. In the given Slater-determinant state, all scattering processes with momentum transfers $p > \bar{p}$ are guaranteed to annihilate the state and thereby satisfy Eq.~\eqref{eq: condition_on_states}. Generically, $\bar{p} < P_{\mathrm{tot}}$ is smaller than the total momentum of the state in question. Hence, a straightforward method to obtain exact eigenstates is to set all potentials $V(1) = V(2) = V(3) = \dots = V(\bar{p})$ equal in order to satisfy Eq.~\eqref{eq: condition_on_potentials} for the remaining allowed scattering processes. 
Such homogeneous potential configurations may be frugal, see for instance the first two rows of Fig.~\ref{fig: exactstates}a. Generically, they also give rise to a large number of exact states at smaller total momenta than $P_{\mathrm{tot}}$ (see App.~\ref{app: exact_tally}). 

We next discuss non-homogeneous potential configurations that generalize to arbitrarily large total momenta. We focus on the simplest types of configurations, for the sake of pedagogy and also because we explicitly study their entanglement entropy in Sec.~\ref{sec: entanglement} (Fig.~\ref{fig: entanglement}d). More intricate exact state solutions, such as those appearing in Fig.~\ref{fig: exactstates}a, are analyzed in Sec.~\ref{sec: selfconsistency}.

\subsection{Alternating-potential sequence} \label{sec: alternatingpotentials}
We begin by considering alternating $V(p)$ configurations, where every second potential $V(1) = V(3) = \dots = V(2n+1)$ is set equal. They give rise to an exact state at total momentum $P_{\mathrm{tot}}=(n+1)(n+2)/2$ with $n\geq0$. (Since $P_{\mathrm{tot}}=20$ is not of this form, there is no alternating $V(p)$ configuration in Fig.~\ref{fig: exactstates}.) Consider the non-interacting ground state $\ket{\Omega}$ in Eq.~\eqref{eq: non-int-groundstate-def}, pictured in Fig.~\ref{fig: exactfrugals}a. We first assume $n\equiv2\bar{k}-1$ is odd. Then, moving the first $\bar{k}$ occupied electrons at odd negative momenta, located at $p = -(2k-1)$, $k=1 \dots \bar{k}$, to the previously unoccupied positive momenta at $-p+1 = 2k$, we obtain the fermionic basis states
\begin{equation} \label{eq: statesexactsequence1}
\ket{\Psi_{\bar{k}}} = \left(\prod_{k'=1}^{\bar{k}} c^\dagger_{2k'} \right) \left(\prod_{k=1}^{\bar{k}} c_{-(2k-1)}\right) \ket{\Omega},
\end{equation}
an example of which is depicted in Fig.~\ref{fig: exactfrugals}b for $\bar{k}=2$. These states have a total momentum
\begin{equation} \label{eq: totalmomexactsequence1}
P_{\mathrm{tot}} = \sum_{k=1}^{\bar{k}} (4k-1) = \bar{k} (2\bar{k}+1) = 3,10,21,\dots,.
\end{equation}
We next assume $n\equiv2(\tilde{k}-1)$ is even. Then, moving the first $\tilde{k}$ occupied electrons at even momenta, located at $p=-2k$, $k=0 \dots \tilde{k}-1$, to the previously unoccupied momenta at $-p+1 = 2k+1$, we obtain the fermionic basis states
\begin{equation} \label{eq: statesexactsequence2}
\ket{\Psi_{\tilde{k}}} = \left( \prod_{k'=0}^{\tilde{k}-1} c^\dagger_{2k'+1} \right) \left(\prod_{k=0}^{\tilde{k}-1} c_{-2k}\right) \ket{\Omega},
\end{equation}
an example of which is depicted in Fig.~\ref{fig: exactfrugals}c for $\tilde{k}=3$. These states have a total momentum 
\begin{equation} \label{eq: totalmomexactsequence2}
P_{\mathrm{tot}} = \sum_{k=0}^{\tilde{k}-1} (4k+1) = \tilde{k}(2\tilde{k}-1) = 1,6,15,\dots,.
\end{equation}
For either type of state, $\ket{\Psi_{\bar{k}}}$ or $\ket{\Psi_{\tilde{k}}}$, scattering to empty sites can only involve odd momentum transfers $p$ in Eq.~\eqref{eq: Hint_destructive_interference}. Hence, when all odd potentials $V(1) = V(3) = \dots = V(2n+1)$ are set equal, where $n=2\bar{k}-1$ or $n=2(\tilde{k}-1)$ depending on whether $n$ is odd or even, respectively, the states $\ket{\Psi_{\bar{k}/\tilde{k}}}$ become exact eigenstates of $\normord{H}$ in Eq.~\eqref{eq: Hint_vanilla}. Indeed, for the exact state pictured in Fig.~\ref{fig: exactfrugals}b, and assuming a potential configuration $V(1) = V(3) = V(5) = V(7)$, either Eq.~\eqref{eq: condition_on_states} or Eq.~\eqref{eq: condition_on_potentials} are fulfilled for all $q,k,p$ in the ranges of Eqs.~\eqref{eq: qkrange} and~\eqref{eq: prange}: for even momentum transfers $p$, Eq.~\eqref{eq: condition_on_states} is satisfied, while for odd momentum transfers $p$, Eq.~\eqref{eq: condition_on_potentials} is satisfied. The same result holds for the exact state pictured in Fig.~\ref{fig: exactfrugals}c when we choose $V(1) = V(3) = V(5) = V(7) = V(9)$.

\subsection{Descendant sequences}
We next show that there are two descendant exact state sequences in all total momentum sectors with $P_{\mathrm{tot}}=2+(n+1)(n+2)/2$ with $n\geq0$. The first descendant series is given by the states
\begin{equation} \label{eq: firstdescendants}
\begin{aligned}
&\ket{\Psi^{(+2,1)}_{\bar{k}}} = c^\dagger_{2\bar{k}+2} \left(\prod_{k'=1}^{\bar{k}-1} c^\dagger_{2k'} \right) \left(\prod_{k=1}^{\bar{k}} c_{-(2k-1)} \right) \ket{\Omega}, \\
&\ket{\Psi^{(+2,1)}_{\tilde{k}}} = c^\dagger_{2\tilde{k}+1} \left(\prod_{k'=0}^{\tilde{k}-2} c^\dagger_{2k'+1}\right) \left(\prod_{k=0}^{\tilde{k}-1} c_{-2k} \right) \ket{\Omega},
\end{aligned}
\end{equation}
which have total momentum $P_{\mathrm{tot}}=2+\bar{k} (2\bar{k}+1)$ and $P_{\mathrm{tot}}=2+\tilde{k}(2\tilde{k}-1)$, respectively. An example is shown in Fig.~\ref{fig: exactfrugals}d. The second descendant series is given by
\begin{equation} \label{eq: seconddescendants}
\begin{aligned}
&\ket{\Psi^{(+2,2)}_{\bar{k}}} = \left(\prod_{k'=1}^{\bar{k}} c^\dagger_{2k'} \right) \left(\prod_{k=1}^{\bar{k}-1} c_{-(2k-1)} \right) c_{-(2\bar{k}+1)} \ket{\Omega}, \\
&\ket{\Psi^{(+2,2)}_{\tilde{k}}} = \left(\prod_{k'=0}^{\tilde{k}-1} c^\dagger_{2k'+1} \right) \left(\prod_{k=0}^{\tilde{k}-2} c_{-2k} \right) c_{-2\tilde{k}} \ket{\Omega},
\end{aligned}
\end{equation}
which have the same total momentum as the first descendants. As an example, see Fig.~\ref{fig: exactfrugals}e. These states are related to the states in Eq.~\eqref{eq: firstdescendants} via the interaction potentials $V(2)$ and $V(2(n+1)+1)$ in Eq.~\eqref{eq: Hint_destructive_interference}, while for all remaining even momentum transfers $p \neq 2$, Eq.~\eqref{eq: condition_on_states} is still satisfied. Correspondingly, if we set the potentials $V(2) = V(1) = V(3) = V(5) = \dots = V(2(n+1)+1)$ equal, both $\ket{\Psi^{(+2,1)}_{\bar{k}/\tilde{k}}}$ and $\ket{\Psi^{(+2,2)}_{\bar{k}/\tilde{k}}}$ become exact eigenstates of $\normord{H}$. Then, for even momentum transfers $p$ except for $p=2$, Eq.~\eqref{eq: condition_on_states} is satisfied, while for odd momentum transfers $p$ as well as $p=2$, Eq.~\eqref{eq: condition_on_potentials} is satisfied. 

The states $\ket{\Psi_{\bar{k}/\tilde{k}}}$ are frugal in all Hilbert space sectors for which we numerically solved Eqs.~\eqref{eq: condition_on_states} and~\eqref{eq: condition_on_potentials} in App.~\ref{app: exact_tally}, while the states $\ket{\Psi^{(+2,1)}_{\bar{k}/\tilde{k}}}$ and $\ket{\Psi^{(+2,2)}_{\bar{k}/\tilde{k}}}$ are frugal starting from $P_\mathrm{tot} = 12$ onwards (but still exact for lower $P_\mathrm{tot}$). In App.~\ref{sec: generalizable_sequences_appendix}, we discuss higher-order descendants of $\ket{\Psi_{\bar{k}/\tilde{k}}}$, which form an infinite family of generically non-frugal exact eigenstates.

\section{General properties of exact eigenstates} \label{sec: genprops}
We here generalize our analytical investigation of the exact eigenstate constraints in Eqs.~\eqref{eq: condition_on_states} and~\eqref{eq: condition_on_potentials} beyond the study of particular solutions. Analyzing the exhaustive set of frugal exact states that was numerically found in App.~\ref{app: exact_tally} for Hilbert space sectors up to $P_{\mathrm{tot}}=40$, two regularities warrant explanation: 

(1) The majority of frugal states, but not all, comes in pairs, in that the same set of potentials $V(p)$ gives rise to two different frugal states (see e.g. rows $1,2; 3,4; 6,7; 8,9; 10,11; 12,13$ in Fig.~\ref{fig: exactstates}a, the exception is row 5 where only one exact state appears). 

(2) The majority of frugal states, but not all, has an electron or hole occupation that has the same structure as their corresponding interaction potential $V(p)$. Specifically, the momenta $p$ for which $V(p)$ must be tuned align with a subset of the occupied electron momenta up to a constant shift, or alternatively with the occupied hole momenta after a reflection $p \rightarrow 1-p$. For instance, the 8 equal potentials $V(p)$ in row $3$ of Fig.~\ref{fig: exactstates}a align with the 8 highest occupied electron momenta of the corresponding exact eigenstate. Conversely, in row 4 of Fig.~\ref{fig: exactstates}a, the same 8 equal potentials align with the 8 lowest occupied hole momenta of a different exact eigenstate that pertains to the same potential configuration. (For more examples, see rows $3,4,8,9,10,11,12,13$ in Fig.~\ref{fig: exactstates}a, the exceptions are rows $1,2,5,6,7$). 

The explanation of these features will provide us with a general method for constructing an infinite number of guaranteed exact eigenstate solutions, which includes all exact state sequences found in Sec.~\ref{sec: generalizable_sequences} as special cases.

\subsection{Particle-hole duality} \label{sec: duality}
The presence of exact state pairs originating from the same potential configuration comes about by a particle-hole duality of the solution space of Eq.~\eqref{eq: condition_on_states}. Paired states are exchanged by the duality transformation, while unpaired states are self-dual.
To derive the duality, we first state the quantities that it leaves invariant. From Eq.~\eqref{eq: totalmomstates}, the total momentum of a fermionic basis state $\ket{\bs{n},\bs{\bar{n}}}$ is given by 
\begin{equation} \label{eq: totmomconst_analytics}
P_\mathrm{tot} = \sum_{p > 0} p \left(n_p + \bar{n}_{-p} \right).
\end{equation}
In addition, our restriction to particle number $\braket{\hat{N}} = 0$ implies 
\begin{equation} \label{eq: totnumconst_analytics}
N_\mathrm{tot} = \sum_{p>0} \left(n_p - \bar{n}_{-p} \right) - \bar{n}_{0} = 0.
\end{equation}
Eqs.~\eqref{eq: totmomconst_analytics} and~\eqref{eq: totnumconst_analytics} specify the Hilbert space sector of $\ket{\bs{n},\bs{\bar{n}}}$. Moreover, it follows from Eq.~\eqref{eq: condition_on_states}, restricted to the ranges of Eqs.~\eqref{eq: qkrange} and~\eqref{eq: prange}, that the interaction component $V(p)$, $p > 0$, must \emph{not} be tuned if
\begin{equation} \label{eq: equivalentconditiononstates}
\bar{n}_{q+p} n_q n_k \bar{n}_{k-p} = \bar{n}_{q-p} n_q n_k \bar{n}_{k+p} = 0
\end{equation}
holds for all $q,k$ in the range of Eq.~\eqref{eq: qkrange}, because this condition guarantees Eq.~\eqref{eq: condition_on_states} for $p$ and $-p$. Here, the second equation follows from noting that $V(p) = V(-p)$.

All of these properties are left invariant under the duality transformation
\begin{equation} \label{eq: duality}
\mathrm{PH}: \quad n_p \leftrightarrow \bar{n}_{-p+1}.
\end{equation}
Specifically, we have
\begin{equation}
\begin{aligned}
\mathrm{PH}: \quad &P_\mathrm{tot} \leftrightarrow P_\mathrm{tot} - N_\mathrm{tot}, \\ &N_\mathrm{tot} \leftrightarrow -N_\mathrm{tot}, \\
& \bar{n}_{q+p} n_q n_k \bar{n}_{k-p} \leftrightarrow \bar{n}_{\tilde{q}-p} n_{\tilde{q}} n_{\tilde{k}} \bar{n}_{\tilde{k}+p}, \\
& \bar{n}_{q-p} n_q n_k \bar{n}_{k+p} \leftrightarrow \bar{n}_{\tilde{q}+p} n_{\tilde{q}} n_{\tilde{k}} \bar{n}_{\tilde{k}-p},
\end{aligned}
\end{equation}
where we have defined $\tilde{q} = 1+p-k$ and $\tilde{k} = 1-p-q$. Since $p>0$ and $q > k$, these satisfy $\tilde{q} > \tilde{k}$ and are thereby compatible with Eq.~\eqref{eq: qkrange}. Hence, if Eqs.~\eqref{eq: totmomconst_analytics}-\eqref{eq: equivalentconditiononstates} hold for $\ket{\bs{n},\bs{\bar{n}}}$, then they also hold for its dual. Correspondingly, if the exact state $\ket{\bs{n},\bs{\bar{n}}}$ is not self-dual (if the duality transformation in Eq.~\eqref{eq: duality} does not recover the same state), then it must have a partner exact state that is induced by the same potential configuration.

We stress that the particle-hole duality in Eq.~\eqref{eq: duality} is \emph{not} a particle-hole symmetry of the Hamiltonian in Eq.~\eqref{eq: Hint_vanilla} itself: for instance, the dispersion $\epsilon(p)$ need not be particle-hole symmetric. Hence, exact eigenstates related by the duality generically lie at different energies: examples are the states in rows 12 and 13 of Fig.~\ref{fig: exactstates}a, whose energy difference in presence of a quadratic kinetic energy perturbation is indicated in Fig.~\ref{fig: exactstates}c. Unlike their energy, however, we shown in Sec.~\ref{sec: entanglement} that the entanglement entropy of duality-related states is identical.

\subsection{Self-consistent exact states} \label{sec: selfconsistency}
For special potential configurations $V(p)$, we can systematically write down a subset of their associated exact eigenstates. We call these \emph{self-consistent} configurations. Given a set of potentials $V(p)$, let $Q^\mathrm{c}$ be the set of momenta $p>0$ where the $V(p)$ are set equal. For any total momentum, we can always assume that $Q^\mathrm{c}$ is finite, because Eq.~\eqref{eq: condition_on_potentials} must only be enforced in the range of Eq.~\eqref{eq: prange} to yield exact states. Then, there is a maximal momentum $\bar{p} = \max (Q^\mathrm{c})$. We define $Q$ as the set of momenta $p>0$ not in $Q^\mathrm{c}$, \emph{i.e.}, the momenta where $V(p)$ can be be chosen freely. Since $Q^\mathrm{c}$ is finite, $Q$ is infinitely large and includes all $p > \bar{p}$. Then, the potential configuration $V(p)$ is self-consistent if and only if
\begin{equation} \label{eq: self_consistency}
\forall \, p_1, p_2 \in Q \quad \rightarrow \quad p_1 + p_2 \in Q,
\end{equation}
that is, $Q$ forms a semigroup under addition. As an example, for the potential configuration in row 3 of Fig.~\ref{fig: exactstates}a, we have $Q^\mathrm{c} = \{1,2,4,5,7,8,10,11\}$ and $Q = \{3,6,9\}\cup\{p \in \mathbb{Z} | p \geq 12\}$. This choice is self-consistent because the elements of $Q$ smaller than $\bar{p} = 11$ exhaust all multiples of $3$ that are smaller than $\bar{p}$. Now, given a momentum $\bar{q}$ with $0 \leq \bar{q} \leq \bar{p}$, we define the two finite sets
\begin{equation}
\begin{aligned}
Q_{\leq \bar{q}} &= \{p \in Q \, | \, 0 < p \leq \bar{q}\}, \\
Q_{>\bar{q}}^{\mathrm{c}} &= \{p \in Q^\mathrm{c} \, | \, \bar{q} < p \leq \bar{p}\}.
\end{aligned}
\end{equation}
For instance, for the previously discussed potential configuration in row 3 of Fig.~\ref{fig: exactstates}a and $\bar{q}=7$, $Q_{\leq 7} = \{3,6\}$ and $Q_{>7}^{\mathrm{c}} = \{8,10,11\}$. In general, for $\bar{q} = 0$, $Q_{\leq 0}$ is empty, and $Q_{>0}^{\mathrm{c}}$ contains all momenta where $V(p)$ is set equal. Successively increasing $\bar{q} \rightarrow \bar{q}+1$ either removes an element from $Q_{>\bar{q}}^{\mathrm{c}}$ (if $\bar{q}+1 \in Q^\mathrm{c}$) or adds an element to $Q_{\leq \bar{q}}$ (if $\bar{q}+1 \in Q$). Finally, for $\bar{q} = \bar{p}$, $Q_{\leq \bar{p}}$ contains all momenta in the range $[1,\bar{p}]$ with $V(p)$ unconstrained, and $Q_{>\bar{p}}^{\mathrm{c}}$ is empty.
Therefore, it is always possible to find a momentum $\bar{q}$ such that 
\begin{equation} \label{eq: setsizecomparison}
|Q_{\leq \bar{q}}| = |Q^{\mathrm{c}}_{>\bar{q}}| - 1,
\end{equation}
where $|X|$ denotes the number of elements in the finite set $X$. As an example, for the previously discussed potential configuration in row 3 of Fig.~\ref{fig: exactstates}a, this amounts to the choice $\bar{q}=7$ because $|Q_{\leq 7}| = 2$ and $|Q_{>7}^{\mathrm{c}}| = 3$. Eq.~\eqref{eq: setsizecomparison} implies that there is one less free potential $V(p)$ in the range $p \leq \bar{q}$ than there are constrained (equal) potentials $V(p)$ in the range $p > \bar{q}$. A guaranteed exact eigenstate associated with the self-consistent potential configuration $V(p)$ is then given by
\begin{equation} \label{eq: selfconsistent_eigenstate}
\ket{\Psi_V} = c_{-\bar{q}} \prod_{p \in Q_{\leq \bar{q}}} c_{p-\bar{q}} \prod_{p \in Q_{>\bar{q}}^{\mathrm{c}}} c^\dagger_{p-\bar{q}} \ket{\Omega},
\end{equation}
which a has total momentum
\begin{equation}
P_V = |\bar{q}| + \sum_{p \in Q_{\leq \bar{q}}} |p-\bar{q}| + \sum_{p \in Q_{>\bar{q}}^{\mathrm{c}}} |p-\bar{q}|.
\end{equation}
As an example, for the previously discussed potential configuration in row 3 of Fig.~\ref{fig: exactstates}a, this state is just the frugal state shown in the same row.

To see that the Slater-determinant state $\ket{\Psi_V}$ is an eigenstate of $\normord{H}$ in Eq.~\eqref{eq: Hint_vanilla} in general, we first list its unoccupied momenta:
\begin{equation} \label{eq: listofholes}
\text{holes}[\ket{\Psi_V}] = \{-\bar{q}\} \cup \{-\bar{q} + p \, | \, p \in Q\}.
\end{equation}
For $p \in Q_{\leq \bar{q}} \subset Q$, the momentum $-\bar{q} + p$ is manifestly unoccupied, which follows from the presence of the annihilation operators $c_{p-\bar{q}}$, one for every $p \in Q_{\leq \bar{q}}$, in the first product of Eq.~\eqref{eq: selfconsistent_eigenstate}. For the remaining $p \in Q$, which must satisfy $p - \bar{q}> 0$, the fact that $-\bar{q} + p$ is unoccupied follows from the absence of creation operators $c^\dagger_{p-\bar{q}}$, $p \in Q$, in the second product of Eq.~\eqref{eq: selfconsistent_eigenstate} -- recall that $\ket{\Omega}$ has all positive momenta unoccupied.

Next, we examine the scattering processes in Eq.~\eqref{eq: Hint_destructive_interference} that move an occupied electron to the lowest-momentum hole of $\ket{\Psi_V}$, located at $-\bar{q}$. These scatterings cannot involve momentum transfers $p \in Q$, because $\ket{\Psi_V}$ has all momenta $-\bar{q}+p$, $p \in Q$ unoccupied. 

Finally, we examine the scattering processes that move an occupied electron to any of the remaining holes of $\ket{\Psi_V}$, which are located at $-\bar{q} + p_1$, $p_1 \in Q$. Due to the self-consistency constraint in Eq.~\eqref{eq: self_consistency}, we have for all momentum transfers $p_2 \in Q$ that 
\begin{equation}
(-\bar{q} + p_1) + p_2 \equiv -\bar{q} + \tilde{p}, \quad \tilde{p} \in Q,
\end{equation}
which, by Eq.~\eqref{eq: listofholes}, is the momentum of another hole.
Therefore, all momenta that can be reached by momentum transfers $p \in Q$ from any hole of $\ket{\Psi_V}$ are guaranteed to be unoccupied, so that scattering cannot occur: for all $p \in Q$, Eq.~\eqref{eq: condition_on_states} is satisfied by $\ket{\Psi_V}$. Moreover, $V(p)$ satisfies Eq.~\eqref{eq: condition_on_potentials} by construction for all $p \notin Q$. Hence, $\ket{\Psi_V}$ is an exact eigenstate of $\normord{H}$ in Eq.~\eqref{eq: Hint_vanilla}. Unless it is invariant under the particle-hole duality of Eq.~\eqref{eq: duality}, another exact eigenstate can be obtained by applying the duality transformation. For example, the state in row 3 of Fig.~\ref{fig: exactstates}a is the particle-hole dual of the state in row 4.

Self-consistent potentials $V(p)$ generically also yield additional exact eigenstates that are not of the form of Eq.~\eqref{eq: selfconsistent_eigenstate}, or related to a state of this form by a particle-hole transformation. For instance, \emph{all} potential configurations in Fig.~\ref{fig: exactstates}a ($P_{\mathrm{tot}}=20$) are self-consistent, however, only rows $3,4,8,9,10,11,12,13$ have states corresponding to Eq.~\eqref{eq: selfconsistent_eigenstate} and its particle-hole dual. The states of the form of Eq.~\eqref{eq: selfconsistent_eigenstate} that are associated with rows $1,2,5,6,7$ of Fig.~\ref{fig: exactstates}a instead appear as non-frugal exact states in sectors with total momentum $P_{V}<20$. However, as is evident from the data presented in App.~\ref{app: exact_tally}, states of the form of Eq.~\eqref{eq: selfconsistent_eigenstate} or their duals form the majority of all frugal states, at least in the Hilbert space sectors up to total momentum $40$. They also include all simple exact state sequences discussed in Sec.~\ref{sec: generalizable_sequences}.

\begin{figure*}[t]
\centering
\includegraphics[width=\textwidth]{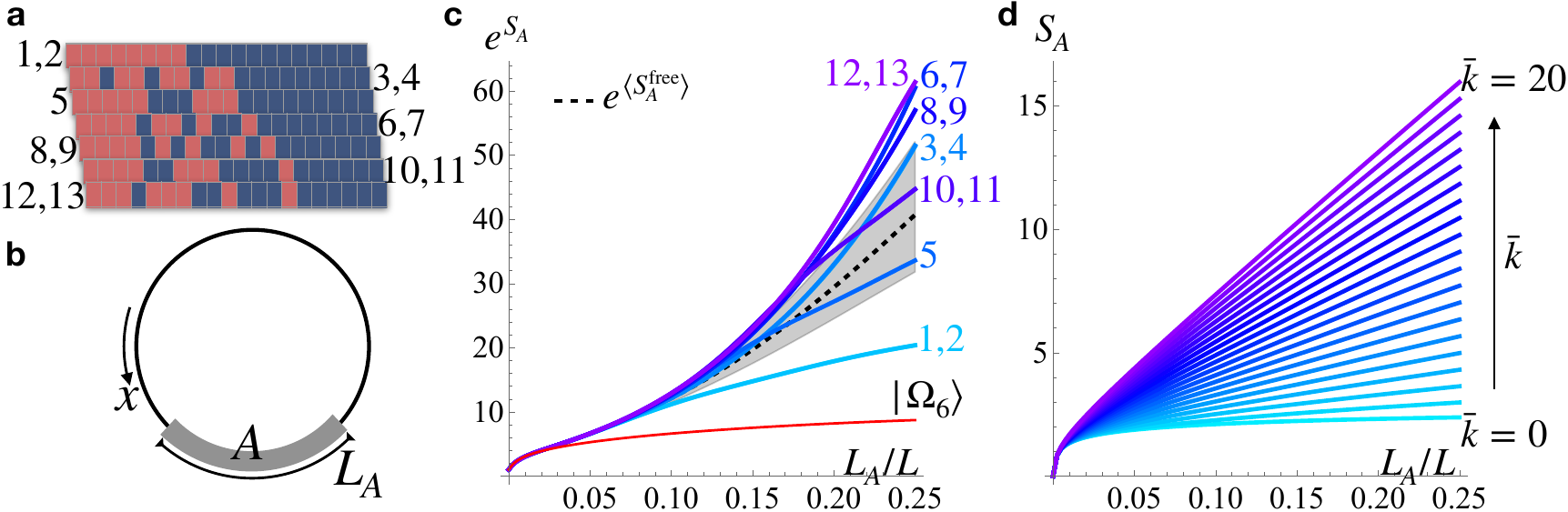}
\caption{Real-space entanglement of exact eigenstates in the $P_\mathrm{tot} = 20$ sector. (a)~Potential tuning patterns giving rise to frugal exact states, where the indices shown indicate the corresponding rows in the right panel of Fig.~\ref{fig: exactstates}a. As explained in the main text, the frugal exact states originating from the same potential configuration are paired by particle-hole duality and have an identical entanglement entropy. (b)~Entanglement cut in a one-dimensional chiral wire with periodic boundary conditions. We calculate the von Neumann entanglement entropy $S_A$ for subsystem $A$ as a function of cut length $L_A$ (for details on how $S_A$ is calculated without lattice discretization, with is unavailable for chiral systems, see App.~\ref{app: real_space_entanglement}). For each exact state, we take the momenta highlighted in the respective line of the right panel of Fig.~\ref{fig: exactstates}a to be occupied, in addition to all lower-lying momenta in the range $[-100, -10]$, implying a momentum cutoff $\Lambda = 100$. We observe that occupying momenta below $p=-100$ shifts the entanglement entropies $S_A$ for $L_A/L \gg 1/100$ by an equal amount and does not affect their asymptotic scaling. This can be understood intuitively by noting that entanglement cuts of length $L_A$ predominantly resolve the momentum scale $p\sim 1/L_A$, so that cuts with $1/L_A \ll \Lambda/L$ are insensitive to occupying additional momenta at the cutoff scale, up to a universal ultraviolet divergence contribution~\cite{WittenEntanglement18}.
(c)~Exponentiated entanglement entropy $e^{S_A}$ as function of cut length for the exact states arising from the potentials in a). An approximately linear scaling of $e^{S_A}$ indicates logarithmic (sub-volume) entanglement growth, while exponential scaling of $e^{S_A}$ implies volume-law growth. For the potentials depicted in a), we find examples of both kinds: the states with labels $1,2$ exhibit a pronounced sub-volume scaling, while the states labelled by $6,7,8,9,12,13$ are compatible with volume-law scaling. The dashed black line and surrounding grey region indicate the exponentiated average entropy $e^{\braket{S_A^{\mathrm{free}}}}$ and its standard deviation over all $\mathcal{P}(20) = 627$ fermionic basis states. Finally, the exponentiated entanglement entropy $e^{S_A [\ket{\Omega_6}]}$ of the Fermi sea state $\ket{\Omega_6}$ (defined in Sec.~\ref{sec: rspaceentanglement}) at total particle number $N=6$, with $P_\mathrm{tot} = 21$, is highlighted in red. (d)~Entanglement entropy $S_A$ of the alternating-potential exact state sequence $\ket{\Psi_{\bar{k}}}$ of Eq.~\eqref{eq: statesexactsequence1} for $\bar{k}=0, \dots, 20$. For the largest $\bar{k}$ we have considered, $P_\mathrm{tot}(\bar{k}= 20) = 820$. We use a momentum cutoff $\Lambda = 200$. The linear scaling of $S_A$ at $\bar{k}$ far from $\bar{k}=0$ ($\ket{\Psi_{\bar{k}=0}}$ corresponds to the filled Fermi sea state $\ket{\Omega}$) implies volume-law entanglement growth.}
\label{fig: entanglement}
\end{figure*}

\section{Entanglement properties} \label{sec: entanglement}
We next investigate the entanglement properties of the exact Slater-determinant eigenstates of the CNLLL. While their real-space entanglement entropy can exhibit a variety of scaling behaviors, ranging from sub-volume to volume-law scaling, their momentum-space entanglement is identically zero. This finding is in stark contrast to the case of typical, non-exact excited eigenstates of the Hamiltonian in Eq.~\eqref{eq: Hint_vanilla}, whose momentum-space entanglement entropy behaves thermally.

Therefore, the exact states of the CNLLL share the core feature of QMBS -- anomalously low entanglement violating the ETH -- but with respect to partitions in momentum space instead of real space. In contrast, generic excited states of the CNLLL are thermalizing, as is confirmed by our level statistics analysis in App.~\ref{app: levelstatistics}.

\subsection{Real space} \label{sec: rspaceentanglement}
Free fermion Slater-determinant states for non-chiral systems are known to exhibit a rich variety of entanglement entropy scaling, ranging from area-law to volume-law~\cite{Lee_2014,Lai15,vafek2017entanglement}. Hence, the real-space entanglement entropy of the exact Slater-determinant eigenstates discussed in Sec.~\ref{sec: exactfermi} is not restricted to be non-thermal, in contrast to many kinds of QMBS that were previously discussed (see reviews~\cite{SerbynReview,PapicReview,SanjayReview}). Indeed, as shown in Fig.~\ref{fig: entanglement} for $P_\mathrm{tot} = 20$ and considering different $V(q)$ distributions that each have $8$ potentials set equal (yielding a total of $13$ different frugal exact states), we find that the exact states can have both sub-volume and volume-law entanglement scaling in real space. Details on how the exact state entanglement entropy is calculated in continuous real space are given in App.~\ref{app: real_space_entanglement}. To cleanly distinguish between the two regimes, we assume that the average entanglement entropy of fermionic basis states $\braket{S_A^{\mathrm{free}}}$, plotted as the black dashed line in Fig.~\ref{fig: entanglement}c, has a volume-law scaling. While this has only been rigorously proven for non-chiral free fermions~\cite{Rigol17}, we expect it to also hold in the chiral case. Since the states with labels $6,7,8,9,12,13$ in Fig.~\ref{fig: entanglement}c have an entanglement entropy that is larger than $\braket{S_A^{\mathrm{free}}}$ by more than a standard deviation, we classify their scaling as volume-law. Similarly, since the states with labels $1,2$ in Fig.~\ref{fig: entanglement}c have an entanglement entropy that is lower than $\braket{S_A^{\mathrm{free}}}$ by more than a standard deviation (even for modest values of $L_A/L$), we classify their scaling as sub-volume. As a reference, we have also included in Fig.~\ref{fig: entanglement}c the entanglement entropy of the filled Fermi-sea state $\ket{\Omega_6} = c^\dagger_1 c^\dagger_2 c^\dagger_3 c^\dagger_4 c^\dagger_5 c^\dagger_6 \ket{\Omega}$. We have chosen the Fermi sea state with $N=6$ because it has a comparable total momentum of $P_\mathrm{tot} = 21$ (in general, the total momentum of $\ket{\Omega_N}$ is given by $N(N+1)/2$) and is known to have a sub-volume entanglement scaling~\cite{Calabrese_2009}. Furthermore, in Fig.~\ref{fig: entanglement}d we study the entanglement entropy of the alternating-potential exact state sequence $\ket{\Psi_{\bar{k}}}$ of Eq.~\eqref{eq: statesexactsequence1} for all $\bar{k}\leq 20$, where $P_\mathrm{tot}(\bar{k}= 20) = 820$. We find that the entropy of the states $\ket{\Psi_{\bar{k}}}$ depends \emph{linearly} on $L_A$ for large $\bar{k}$, with a slope proportional to $\bar{k}$, implying volume-law scaling.

Although the particle-hole duality in Eq.~\eqref{eq: duality} does not in general leave the energy of a fermionic basis state $\ket{\bs{n},\bs{\bar{n}}}$ invariant, it does preserve its real-space entanglement entropy. This is because, as shown in App.~\ref{app: real_space_entanglement}, the entanglement entropy of any given state is invariant under both momentum reflection (effecting $p \rightarrow 1-p$) and particle hole conjugation (effecting $n_p \rightarrow \bar{n}_p$). This symmetry is only exact in the limit where the momentum cutoff $\Lambda$ is sent to infinity, whereas in numerical calculations we must choose a finite cutoff. Nevertheless, in Fig.~\ref{fig: entanglement}c where $\Lambda = 100$, the entanglement entropies of different exact states that are associated with the same potential configuration $V(p)$, and related by particle-hole duality, are indistinguishable by the eye.

In App.~\ref{sec: naive_SA_estimate}, we analytically derive a naive estimate of the maximal possible real-space entanglement entropy $S^{\mathrm{max}}_A$ for a subregion $A$, based on Hilbert space dimension. Unfortunately, this estimate is much larger than the entanglement entropy $S_A$ of \emph{any} exact state at fixed total momentum $P_\mathrm{tot}$, or any other fermionic basis state. The discrepancy stems from our difficulty in taking into account the effects of chirality and total momentum conservation when estimating $S^{\mathrm{max}}_A$ (see App.~\ref{sec: naive_SA_estimate} for details). This prevents us from quantitatively assessing how similar the exact states are to thermal states, which have near-maximal entanglement. Nevertheless, we argue that the qualitative difference between volume and sub-volume scaling evident in Fig.~\ref{fig: entanglement} indicates a broad range of possible real-space entanglement scaling.

\subsection{Momentum space} \label{sec: momspace_ent}
\begin{figure}[t]
\centering
\includegraphics[width=0.48\textwidth]{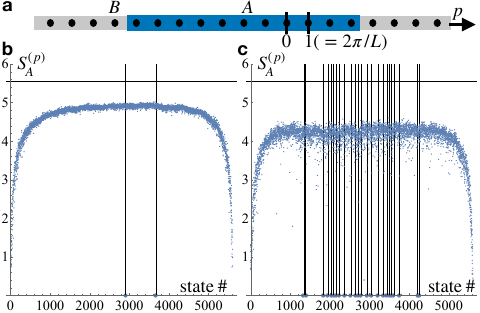}
\caption{Momentum-space entanglement. (a)~We subdivide momentum space into two subregions, $A$ and $B$, that have approximately equal Hilbert space dimensions. Here, for $P_{\mathrm{tot}} = 30$, which has a total dimension of $\mathcal{P}(30) = 5604$, we choose $A = [-7,3]$, and $B$ as its complement, yielding Hilbert space dimensions $\mathrm{dim}(A) = 1086$ and $\mathrm{dim}(B) = 996$. (b)~Entanglement entropy calculated from the eigenvalues of the reduced density matrix in subregion $A$ for a system with two exact states, which have exactly zero momentum-space entanglement. The horizontal line indicates the Page estimate~\cite{Page93,Bianchi19,Murthy19} of the average entanglement entropy. We have set $\epsilon(p) = vp + ap^2$ with $v = 1$ and $a = 0.1/P_\mathrm{tot}$, and have chosen $V(q) = 0.1$ for the $9$ momenta indicated in red in the third row of the $P_\mathrm{tot} = 30$ panel of Fig.~\ref{fig: frugalstates} in App.~\ref{app: exact_tally}, while for all remaining momenta $V(q)$ is sampled uniformly from $[0.05,0.15]$. (c)~Entanglement entropy $S_A^{(p)}$ for a system with $25$ exact states, which follow from setting $V(q) = 0.1$ for $q\leq 11$, while for all remaining momenta $V(q)$ is sampled uniformly from $[0.05,0.15]$ (other parameters left unchanged).}
\label{fig: momspaceentanglement}
\end{figure}

Since the exact eigenstates of the CNLLL are Slater-determinant states built from single-particle states labelled by a momentum quantum number $p$, their entanglement entropy for decompositions in momentum space~\cite{Thomale10,Mondragon13,Lundgren14,Andrade_2014} (Fig.~\ref{fig: momspaceentanglement}a) is identically zero. On the other hand, the remaining correlated eigenstates of $\normord{H}$ in Eq.~\eqref{eq: Hint_vanilla} generically have non-vanishing momentum space entanglement, as shown in Fig.~\ref{fig: momspaceentanglement}b for the case $P_\mathrm{tot} = 30$. In fact, as is evident from that figure, for a $V(q)$ configuration that is constrained to yield two frugal exact states, most eigenstates of $\normord{H}$ are maximally mixed in momentum space: their momentum-space entanglement entropy is numerically comparable to the average entropy of randomly sampled states in the same Hilbert space; \emph{i.e.}, the Page estimate~\cite{Page93}, as indicated by the horizontal line in Fig.~\ref{fig: momspaceentanglement}b,c. Now, as more potentials $V(q)$ are set equal, giving rise to a larger number of exact states with zero momentum-space entanglement, the entropy of the remaining states is only slightly lowered (Fig.~\ref{fig: momspaceentanglement}c).

We stress that in momentum space, the Page estimate applies, as opposed to the real-space cut discussed previously. This is because, for an entanglement cut separating subregions $A$ and $B$ in momentum space, the Hilbert space $\mathcal{H}(P_{\mathrm{tot}})$ at total momentum $P_{\mathrm{tot}}$ factorizes after resolving the total subregion momentum~\cite{Thomale10,Lundgren14}: 
\begin{equation}
\mathcal{H}(P_{\mathrm{tot}}) = \bigoplus_{P_A=0}^{P_{\mathrm{tot}}} \left[\mathcal{H}_A(P_A) \otimes \mathcal{H}_B(P_{\mathrm{tot}}-P_A) \right],
\end{equation}
where $\mathcal{H}_\alpha(P_\alpha)$ is the Hilbert space of momentum-space subregion $\alpha=A,B$ at total subregion momentum $P_\alpha$.
This decomposition property is required to compute the Page estimate in presence of total momentum conservation~\cite{Bianchi19,Murthy19}.
In contrast, the Hilbert space of chiral fermions does not decompose as a tensor product in real space~\cite{hellerman2021quantum}, which is directly related to the impossibility of a lattice discretization for chiral fermions~\cite{NIELSEN1981219}.

To our knowledge, the global thermal behavior of this interacting many-body quantum system was never before studied in momentum space. Our findings demonstrate that the exact states discussed here share the core properties of QMBS -- in being low-entanglement excited states embedded in a continuum of highly-entangled states -- when analyzed with respect to their momentum-space structure.

\section{Summary and discussion} \label{sec: discussion}
We have shown that chirality and fermionic exchange statistics imply the existence of exact eigenstates for a large family of density-density interaction potentials in the CNLLL. Since these eigenstates are of Slater-determinant form, they remain exact eigenstates for any choice of uni-directional single-particle dispersion relation, including non-linearities of arbitrary order. Their existence presupposes certain constraints on the Fourier components $V(p)$ of the interaction potential, in that a subset of these components needs to be of equal strength. It is important to note that this poses only light constraints when considering realistic systems: short-range interactions, which are ubiquitous in metals due to screening effects, are highly localized in position space. Therefore, their Fourier transform depends only weakly on momentum. The complete set of interaction potentials giving rise to exact states of the form discussed here could thus contain a large subset of physically relevant short-range interactions.

We note that the scar states uncovered here have further non-thermal features that go beyond their anomalous (zero) momentum space entanglement entropy. Indeed, since these states are simultaneous eigenstates of all single-particle number operators $\hat{n}_p$, the expectation value $\braket{\hat{n}_p}$ only assumes discrete values $\braket{\hat{n}_p}=0,1$. On the other hand, $\braket{\hat{n}_p}$ follows a continuous Fermi-Dirac distribution in states satisfying the ETH~\cite{rigol2008thermalization}. Similarly, the spectrum of the scar state two-point correlation matrix $\rho_{xy}=\braket{c^\dagger_x c_y}$ exhibits a sharp discontinuity between two discrete eigenvalues $1$ and $0$, which is absent for thermal states~\cite{Bera15}.

We recall that scar states with equal energy level spacings are usually not product states in either real or momentum space (they are obtained by acting with raising or lowering operators on such product states)~\cite{vafek2017entanglement,BerislavSGA19,moudgalya2020eta,mark2020unified,ren2020quasisymmetry,odea2020from}. Nevertheless, their equidistant energy spectrum gives rise to periodic revivals in the time evolution of product states. Conversely, the scar states in the CNLLL -- which do \emph{not} have equal energy spacing -- are (momentum-space) product states by themselves.

Finally, we note that while we have argued at the end of Sec.~\ref{sec: exactfermi} that a uni-directional dispersion relation is essential to stabilize exact eigenstates, our approach can be generalized to multiple species of co-propagating chiral fermions.

\begin{acknowledgments}
We thank Beno\^it Estienne, Chaitanya Murthy, and Vir B. Bulchandani for helpful discussions. F.S. was supported by a fellowship at the Princeton Center for Theoretical Science. This work is part of a project that has received funding from the European Research Council (ERC) under the European Union’s Horizon 2020 research and innovation programme (grant agreement No. 101020833). This work also received support from the DOE Grant No. DE-SC0016239, the Schmidt Fund for Innovative Research, Simons Investigator Grant No. 404513, the Packard Foundation, the Gordon and Betty Moore Foundation through Grant No. GBMF8685 towards the Princeton theory program, and a Guggenheim Fellowship from the John Simon Guggenheim Memorial Foundation. Further support was provided by the NSF-EAGER Grant No. DMR 1643312, NSF-MRSEC Grant No. DMR-1420541 and DMR-2011750, ONR Grant No. N00014-20-1-2303, BSF Israel US foundation Grant No. 2018226, and the Princeton Global Network Funds.

During the long preparation of this work, QMBS in the CNLLL were also reported in Ref.~\onlinecite{martin2021scar}, which discusses the persistence of bosonic or fermionic states as \emph{approximate} eigenstates away from integrability, assuming a Coulomb interaction. Conversely, here we have shown that fermionic states can persist as \emph{exact} eigenstates of the CNLLL, assuming a fine-tuned interaction.
\end{acknowledgments}

\bibliography{references}

\widetext
\appendix

\section{Derivation of Eq.~\eqref{eq: Hint_destructive_interference}} \label{app: Hint_derivation}
In order to derive Eq.~\eqref{eq: Hint_destructive_interference}, we start with Eq.~\eqref{eq: Hint_vanilla}:
\begin{equation} \label{eq: massageinteraction1}
\begin{aligned}
\normord{H_\mathrm{int}} =& \sum_{p>0} V(p) \sum_{qk} c^\dagger_{q+p} c_q c^\dagger_{k-p}c_k \\=& \sum_{p>0} V(p) \sum_{q>k} \left(c^\dagger_{q+p} c_q c^\dagger_{k-p}c_k 
+ c^\dagger_{k+p} c_k c^\dagger_{q-p}c_q \right) 
\\=& \sum_{p>0} \sum_{q>k} \bigg[V(p) c^\dagger_{q+p} c_q c^\dagger_{k-p}c_k + V(q-k+p) c^\dagger_{q+p} c_k c^\dagger_{k-p}c_q \bigg] + \sum_{q>k} \sum_{0 < p \leq q-k} V(p) c^\dagger_{k+p} c_k c^\dagger_{q-p}c_q.
\end{aligned}
\end{equation}
In the second line, we have used that the term with $q=k$ vanishes because $p>0$, allowing us to decompose the sum $\sum_{qk} = \sum_{q>k}+\sum_{k>q}$, and have then exchanged the labels $q \leftrightarrow k$ in the second sum. To obtain the third line, we have substituted $p \rightarrow q-k+p$ only when $p > q-k$ (which is non-negative) in the second term of the second line. For the remaining cases where $p \leq q-k$, we keep the second term of the second line unchanged, giving rise to the third term in the third line. We continue to manipulate
\begin{equation}
\begin{aligned}
\normord{H_\mathrm{int}} =& \sum_{p>0} \sum_{q>k} \left[V(p) - V(q-k+p) \right] c^\dagger_{q+p} c_q c^\dagger_{k-p}c_k + \sum_{q>k} \left(\sum_{0 < p < (q-k)/2} + \sum_{(q-k)/2 < p \leq q-k}\right) V(p) c^\dagger_{k+p} c_k c^\dagger_{q-p}c_q
\\=& \sum_{p>0} \sum_{q>k} \left[V(p) - V(q-k+p) \right] c^\dagger_{q+p} c_q c^\dagger_{k-p}c_k + \sum_{q>k} V(q-k) c^\dagger_{q} c_k c^\dagger_{k}c_q
\\&+ \sum_{q>k} \,\, \sum_{0 < p < (q-k)/2} \left[V(p) c^\dagger_{k+p} c_k c^\dagger_{q-p}c_q + V(q-k-p) c^\dagger_{q-p} c_k c^\dagger_{k+p}c_q \right],
\end{aligned}
\end{equation}
where we have first further split up the second sum in Eq.~\eqref{eq: massageinteraction1} into momenta $p$ above and below $(q-k)/2$. Note that even if $(q-k)/2$ is an integer and thereby a valid momentum (recall that $2\pi/L=1$), the term where $p=(q-k)/2$ does not contribute to the sum due to fermionic statistics. Then, in the second line, we have singled out the $p=0$ contribution and substituted $p \rightarrow q-k-p$ in the sum ranging over $(q-k)/2 < p \leq q-k$. Finally, we simplify:
\begin{equation}
\begin{aligned}
\normord{H_\mathrm{int}} =& \sum_{p>0} \sum_{q>k} \left[V(p) - V(q-k+p) \right] c^\dagger_{q+p} c_q c^\dagger_{k-p}c_k + \sum_{q>k} V(q-k) c^\dagger_{q} c_k c^\dagger_{k}c_q
\\&+ \sum_{q>k} \sum_{0 > p > (k-q)/2} \left[V(-p) - V(q-k+p) \right] c^\dagger_{q+p} c_q c^\dagger_{k-p} c_k
\\=& \sum_{q>k} \left[V(q-k) c^\dagger_{q} c_q c_k c^\dagger_{k} + \sum_{\substack{p \neq 0\\p > (k-q)/2}} \left[V(q-k+p) - V(|p|)\right] c^\dagger_{q+p} c_q c_k c^\dagger_{k-p} \right]
\\=& \sum_{q>k} \sum_{p > (k-q)/2} \left[V(q-k+p) - \delta_{p \neq 0} V(|p|)\right] c^\dagger_{q+p} c_q c_k c^\dagger_{k-p}.
\end{aligned}
\end{equation}
Here, we have substituted $p \rightarrow -p$ in the last sum of the first line, and then introduced 
\begin{equation}
\delta_{p \neq 0} = 
\begin{cases}
1 \quad p \neq 0, \\
0 \quad \text{else},
\end{cases}
\end{equation}
to also absorb the $p=0$ contribution into a single term. We have thus derived Eq.~\eqref{eq: Hint_destructive_interference}.

\begin{figure}[t]
\centering
\includegraphics[width=0.82\textwidth]{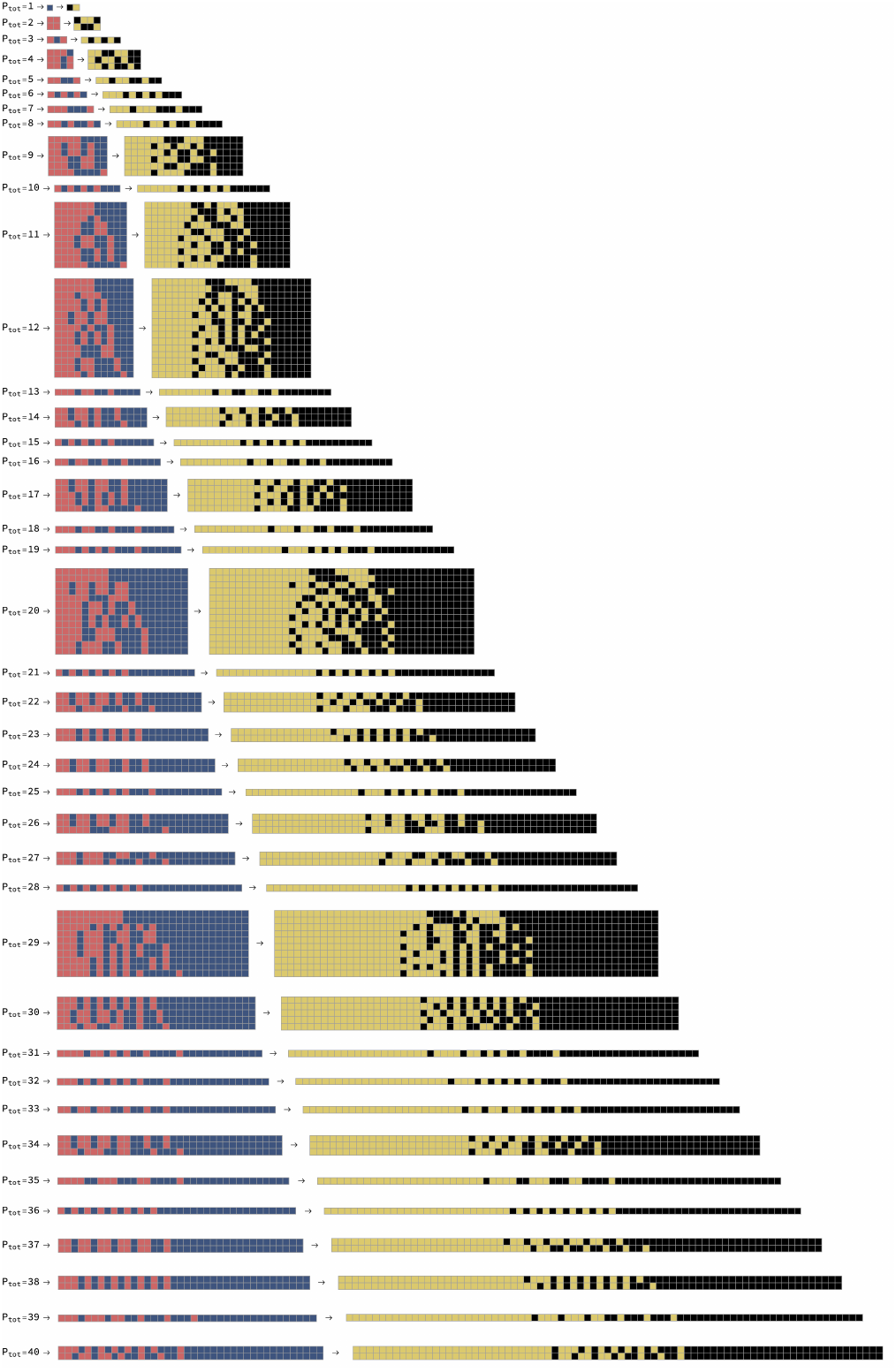}
\caption{Frugal exact state overview. We use the same conventions as in Fig.~\ref{fig: exactstates}a, except that the right panel shows the full momentum range $[-P_{\mathrm{tot}}+1,P_{\mathrm{tot}}]$. For each $P_{\mathrm{tot}}$, we show: (1)~The value of $P_{\mathrm{tot}}$. (2)~The minimal $V(p)$ tuning patterns giving rise to exact states. All $V(p)$ where $p$ is indicated in red need to be set equal for the states shown in 3) to be exact. Tuning a smaller number of potentials does not give rise to exact states. (3)~Frugal states resulting from the respective tuning patterns. These are Slater-determinant states where the electronic modes at momenta indicated in yellow are occupied. All modes at momenta $p>P_{\mathrm{tot}}$ or $p<1-P_{\mathrm{tot}}$ are fully empty or fully occupied, respectively, and are therefore not shown.}
\label{fig: frugalstates}
\end{figure}

\begin{figure}[t]
\centering
\includegraphics[width=0.86\textwidth]{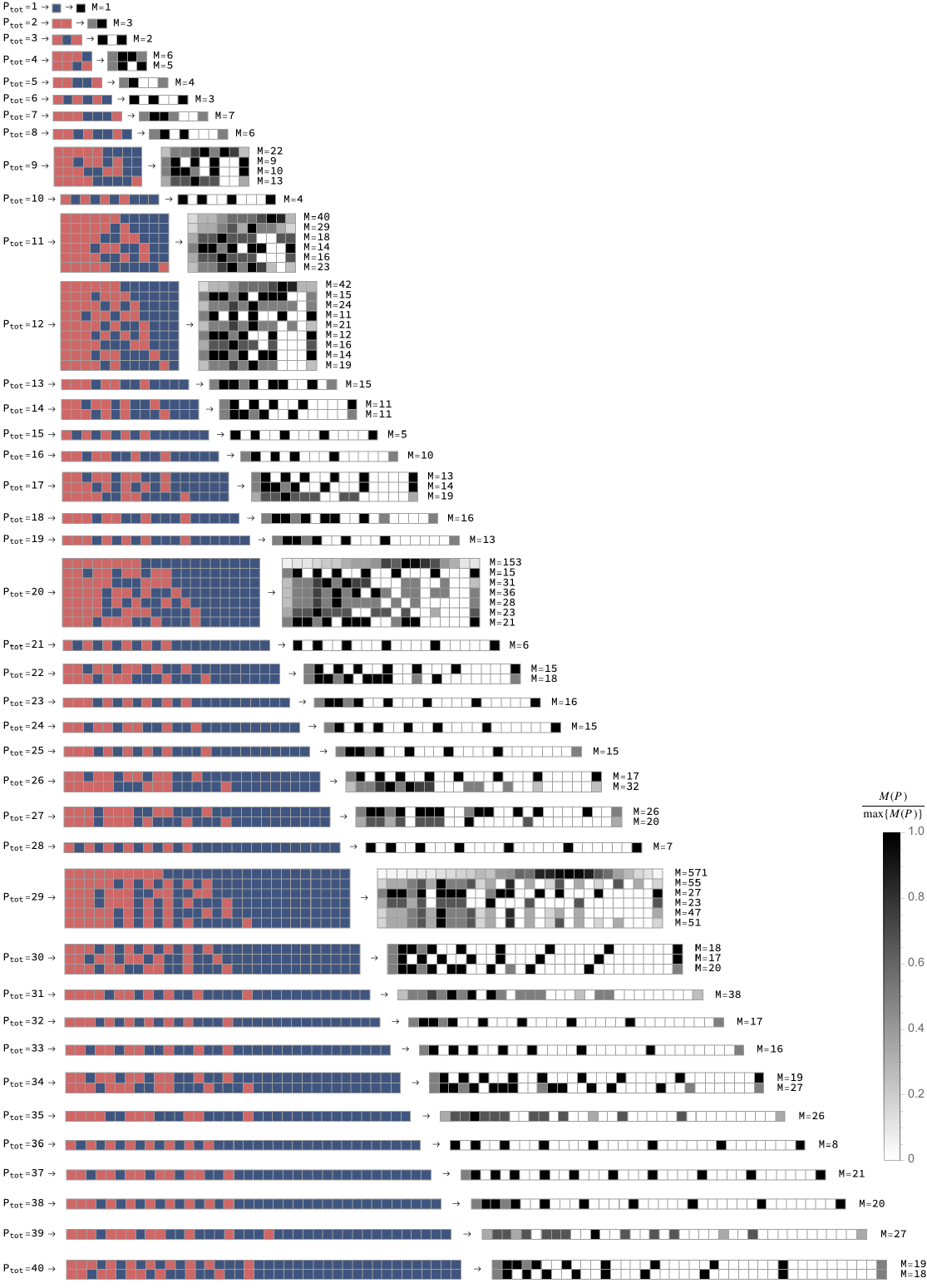}
\caption{Frugal state potential tuning patterns and symmetry sectors containing exact states as a consequence. We use the same conventions as in Fig.~\ref{fig: exactstates}a. For each total momentum $P_{\mathrm{tot}}$, we show: (1)~The value of $P_{\mathrm{tot}}$. (2)~The $V(q)$ tuning patterns giving rise to frugal exact states, also shown in Fig.~\ref{fig: frugalstates}. (3)~The list of total momentum sectors $P$ containing exact states as a result of the given $V(p)$ configuration. Each box corresponds to a choice of $P$, ranging from $P=1$ to $P=P_{\mathrm{tot}}$. The gray scale indicates the ratio $M(P)/\max(\{M(P) | 1 \leq P \leq P_{\mathrm{tot}}\})$. The list of participating sectors contains $P=P_{\mathrm{tot}}$ as its largest element, but also captures (potentially non-frugal) exact states in sectors with smaller total momentum $P<P_{\mathrm{tot}}$ that arise from truncating $V(q)$ at $q=P$. (4)~The total number $M = \sum_{P=1}^{P_{\mathrm{tot}}} M(P)$ of resulting exact states across all $P_{\mathrm{tot}}$ sectors.}
\label{fig: frugalpower}
\end{figure}

\section{All frugal exact states up to $P_{\mathrm{tot}} = 40$} \label{app: exact_tally}
In Fig.~\ref{fig: frugalstates}, we list the frugal exact states in all Hilbert space sectors up to $P_{\mathrm{tot}} = 40$. We use the same conventions as in Fig.~\ref{fig: exactstates}a, except that the right panel shows the full momentum range $[-P_{\mathrm{tot}}+1,P_{\mathrm{tot}}]$. For a given choice of $V(p)$ and a given total momentum $P_{\mathrm{tot}}$ we find either $1$ or $2$ frugal states. Note that while Figs.~\ref{fig: exactstates}a and~\ref{fig: frugalstates} show more than $2$ frugal states for some $P_{\mathrm{tot}}$ sectors, there are at most $2$ frugal states \emph{per choice of} $V(p)$: hence, in some sectors, there are multiple inequivalent frugal $V(p)$ configurations, with each configuration contributing at most $2$ frugal states \emph{in the same sector}.

Moreover, a given frugal choice of $V(p)$ also gives rise to further, not necessarily frugal, exact states in sectors with lower total momentum. For instance, the $V(p)$ configuration shown in the first row of Fig.~\ref{fig: exactstates}a, which has $8$ potentials set equal and is frugal for $P_{\mathrm{tot}}=20$, also gives rise to exact states at $P_{\mathrm{tot}}=19$. These do not show up in the $P_{\mathrm{tot}}=19$ panel of Fig.~\ref{fig: frugalstates} because they are not frugal: the single frugal state at $P_{\mathrm{tot}}=19$ requires $7$ instead of $8$ equal potentials. In Fig.~\ref{fig: frugalpower}, we show all $P_{\mathrm{tot}}$ sectors containing exact states as a result of tuning $V(p)$ to one of the frugal patterns shown in Fig.~\ref{fig: frugalstates}, as well as the total number of resulting exact states \emph{across all total momentum sectors}. 

Given a choice of $V(p)$ that is frugal in the sector with total momentum $P_{\mathrm{tot}}$, let $M(P)$ denote the number of resulting exact states at total momentum $P$, where $1\leq P \leq P_{\mathrm{tot}}$. While for $P = P_{\mathrm{tot}}$, $M(P_{\mathrm{tot}}) \in \{1,2\}$ as noted previously, for $P \leq P_{\mathrm{tot}}$, $M(P)$ can be much larger. Based on Fig.~\ref{fig: frugalpower}, we make several observations: 

(1)~Homogeneous $V(p)$ configurations that have a whole range of potentials $V(1) = V(2) = \dots = V(n)$ set equal lead to smooth $M(P)$ distributions that assume a maximum at $\bar{P}$, where $n < \bar{P} < P_{\mathrm{tot}}$. Naively, one might expect that $\bar{P} = n$, which is the largest Hilbert space sector that has \emph{all} fermionic basis states rendered exact by setting $V(1) = V(2) = \dots = V(n)$. However, the smaller fraction of exact states in sectors at $P>n$ is balanced by the larger Hilbert space dimension $\mathcal{P}(P)$ of these sectors (recall that $\mathcal{P}(x)$ counts the integer partitions of $x$). As an example, see the first rows of $P_{\mathrm{tot}} = 20$ and $29$, respectively, in Fig.~\ref{fig: frugalpower}. 

(2)~Alternating $V(p)$ configurations that have every second potential $V(1) = V(3) = \dots = V(2n+1)$ set equal lead to a single frugal state in all total momentum sectors with $P=(m+1)(m+2)/2$, $m=0 \dots n$, and no further non-frugal exact states in other sectors. These potential distributions were discussed in Sec.~\ref{sec: alternatingpotentials} of the main text. As an example, see $P_{\mathrm{tot}} = 21$ and $28$ in Fig.~\ref{fig: frugalpower}, which respectively realize the case $n=5$ and $n=6$. For instance, the alternating $V(p)$ configuration that is frugal at $P_{\mathrm{tot}} = 21$ also renders the five frugal states at $P=3,6,10,15$ exact (Fig.~\ref{fig: frugalstates}), in addition to the trivial exact state at $P=1$. This is because the choice $V(1) = V(3) = \dots = V(2n+1)$ implies the choice $V(1) = V(3) = \dots = V(2(n-1)+1)$. Therefore, we obtain $M = \sum_{P=1}^{21} M(P) = 6$.

(3)~Intermediate $V(p)$ configurations that consist of disconnected islands of equal potentials have an $M(P)$ distribution that interpolates between the two cases (1) and (2). See for example $P_{\mathrm{tot}} = 31$ and $35$ in Fig.~\ref{fig: frugalpower}.

\section{Level statistics analysis} \label{app: levelstatistics}

\begin{figure}[t]
\centering
\includegraphics[width=1\textwidth]{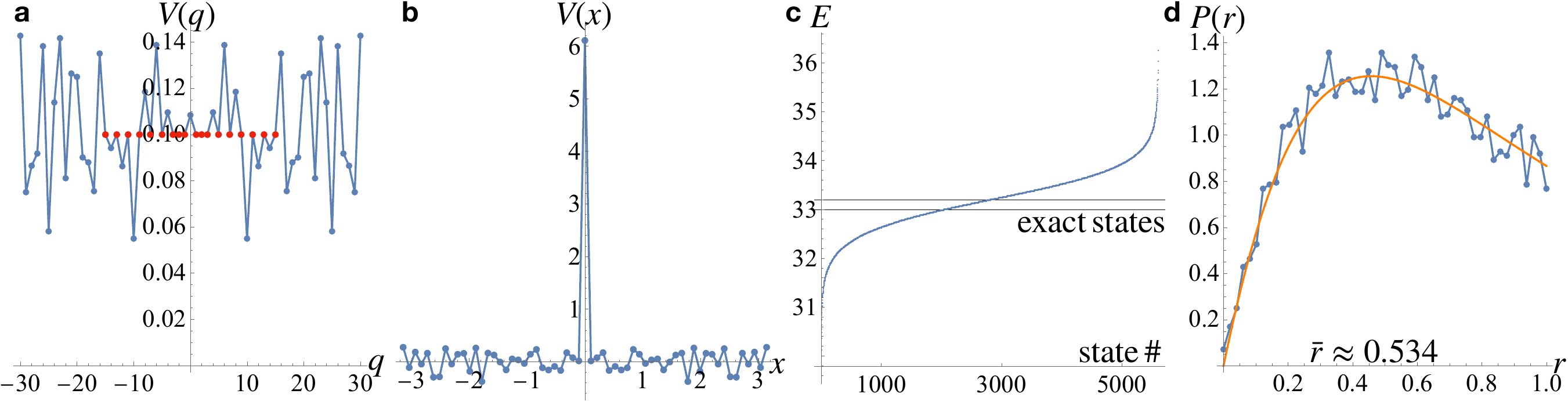}
\caption{Level statistics for a maximally random pattern of potentials $V(p)$ that allows for a non-zero number of exact states in the Hilbert space sector with $P_{\mathrm{tot}} = 30$. We have set $L=2\pi$ for convenience. (a)~To ensure the presence of exact eigenstates, we set to $0.1$ all potentials highlighted in red in the first row of the $P_{\mathrm{tot}} = 30$ panel (left-hand side) of Fig.~\ref{fig: frugalstates} (indicated in red). The remaining potentials are drawn uniformly from the range $[0.05,0.15]$. (b)~Since the momentum space structure of $V(p)$ is non-analytic, the resulting real-space potential $V(x)$ does not decay with distance -- although it is prominently peaked at $x=0$, it features small disorder contributions at finite $x$. (c)~Full energy spectrum of $\normord{H}$ in Eq.~\eqref{eq: Hint_vanilla}. We have set $\epsilon(p) = vp + ap^2$ with $v = 1$ and $a = 0.1/P_\mathrm{tot}$. The energies of the two exact states, corresponding to the first two rows of the $P_{\mathrm{tot}} = 30$ panel (right-hand side) of Fig.~\ref{fig: frugalstates}, are indicated by horizontal bars. (d)~Probability of the adjacent gap ratio $r$, defined in Eq.~\eqref{eq: adjacentgap}. The orange line shows the exact result for the Gaussian orthogonal random matrix ensemble~\cite{Atas13}. We obtain an average adjacent gap ratio of $\bar{r}\approx 0.534$.}
\label{fig: lvlstat1}
\end{figure}

\begin{figure}[t]
\centering
\includegraphics[width=1\textwidth]{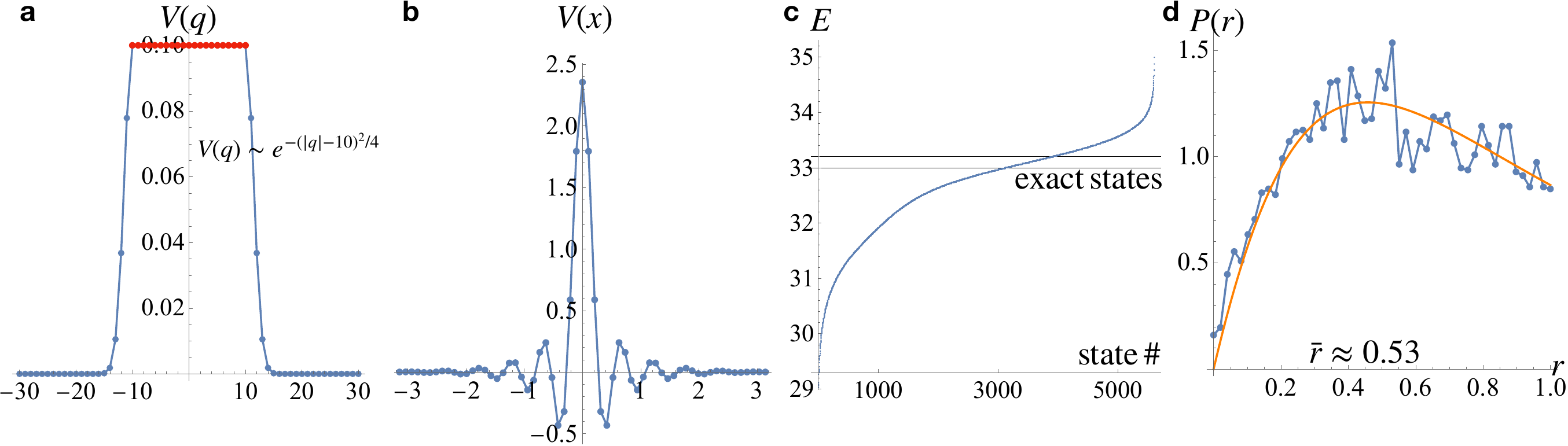}
\caption{Level statistics for a non-random pattern of potentials $V(p)$, corresponding to a slowly exponentially-decaying interaction in real space, that gives rise to two exact states in the Hilbert space sector with $P_{\mathrm{tot}} = 30$. We have set $L=2\pi$ for convenience. (a)~To ensure the presence of exact eigenstates we set to $0.1$ all potentials $V(p)$ with $|p|\leq 10$ (indicated in red). For the remaining momenta, $V(p)$ decays as a Gaussian, with its functional form indicated in the figure. (b)~Since the momentum space structure of $V(p)$ is continuous, the resulting real-space potential $V(x)$ decays only slowly as a result of the abrupt drop in $V(p)$. Since the decay of $V(x)$ is slow, however, $V(x)$ is not well approximated by a delta-function or strictly local potential. (c)~Full energy spectrum of $\normord{H}$ in Eq.~\eqref{eq: Hint_vanilla}. We have set $\epsilon(p) = vp + ap^2$ with $v = 1$ and $a = 0.1/P_\mathrm{tot}$. The energies of the two exact states are indicated by horizontal bars. (d)~Probability of the adjacent gap ratio $r$, defined in Eq.~\eqref{eq: adjacentgap}. The orange line shows the exact result for the Gaussian orthogonal random matrix ensemble~\cite{Atas13}. We obtain an average adjacent gap ratio of $\bar{r}\approx 0.53$.}
\label{fig: lvlstat2}
\end{figure}

\begin{figure}[t]
\centering
\includegraphics[width=1\textwidth]{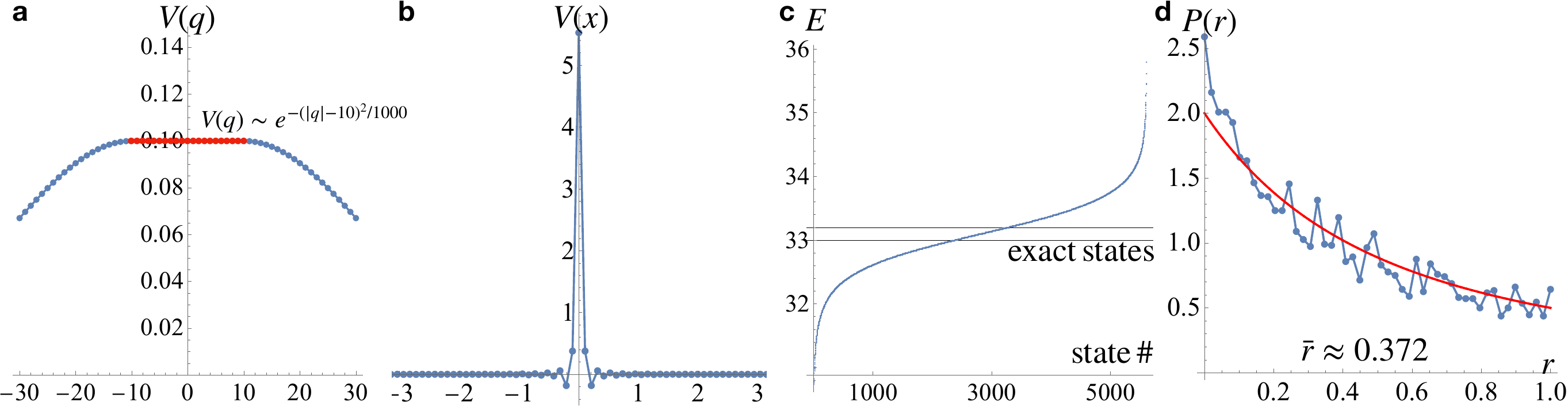}
\caption{Level statistics for a non-random pattern of potentials $V(p)$, corresponding to a short-range interaction in real space, that gives rise to two exact states in the Hilbert space sector with $P_{\mathrm{tot}} = 30$. We have set $L=2\pi$ for convenience. (a)~To ensure the presence of exact eigenstates we set to $0.1$ all potentials $V(p)$ with $|p|\leq 10$ (indicated in red). For the remaining momenta, $V(p)$ decays as a Gaussian, with its functional form indicated in the figure. (b)~Since the momentum space structure of $V(p)$ is smooth, the resulting real-space potential $V(x)$ is short-range. (c)~Full energy spectrum of $\normord{H}$ in Eq.~\eqref{eq: Hint_vanilla}. We have set $\epsilon(p) = vp + ap^2$ with $v = 1$ and $a = 0.1/P_\mathrm{tot}$. The energies of the two exact states are indicated by horizontal bars. (d)~Probability of the adjacent gap ratio $r$, defined in Eq.~\eqref{eq: adjacentgap}. The red line shows the result for integrable systems satisfying Poisson level statistics~\cite{Atas13}. We obtain an average adjacent gap ratio of $\bar{r}\approx 0.372$.}
\label{fig: lvlstat3}
\end{figure}

To access thermalization properties, we study the energy level statistics of $\normord{H}$ in Eq.~\eqref{eq: Hint_vanilla} for various potential distributions containing exact states, while fixing the dispersion to $\epsilon(p) = vp + ap^2$ with $v=1$ and $a = 0.1/P_\mathrm{tot}$. More precisely, we use the adjacent gap ratio, introduced in Ref.~\cite{Oganesyan07}:
\begin{equation} \label{eq: adjacentgap}
r_n = \frac{\min(s_n,s_{n-1})}{\max(s_n,s_{n-1})}, \quad s_n = e_{n+1}-e_n,
\end{equation}
where $e_n$ are the sorted eigenvalues of the Hamiltonian ($e_{n+1} \geq e_n$). The probability distribution $P(r)$ is known for integrable systems, where it follows Poisson level statistics, and various random matrix ensembles, where it follows Wigner-Dyson level statistics~\cite{Atas13}. In particular, integrable systems exhibit an average adjacent gap ratio $\bar{r}\approx 0.386$, while matrices drawn from the Gaussian orthogonal ensemble (GOE) exhibit $\bar{r}\approx 0.536$. Noting that $\normord{H}$ is purely real, GOE is the appropriate ensemble for our purposes, as it corresponds to a uniform distribution over symmetric real matrices. Comparing the $P(r)$ of $\normord{H}$ in various parameter regimes with these known cases then allows us to deduce to what extent the system is integrable or ergodic. 

We study the level statistics of $\normord{H}$ with a quadratic perturbation, $\epsilon(p) = vp + ap^2$, for different interactions $V(p)$ in Figs.~\ref{fig: lvlstat1}-\ref{fig: lvlstat3}. Fig.~\ref{fig: lvlstat1} shows the spectrum and level statistics for a potential $V(x)$ that is sharply peaked at $x=0$ but has long-range disorder. The resulting level statistics is of Wigner-Dyson type, indicating thermalization. Fig.~\ref{fig: lvlstat2} shows the spectrum and level statistics for a potential $V(x)$ that is peaked at $x=0$ but slowly decaying. The resulting level statistics is also of Wigner-Dyson type. Finally, Fig.~\ref{fig: lvlstat3} shows the spectrum and level statistics for a potential $V(x)$ that is peaked at $x=0$, and decaying very fast away from its maximum. The resulting level statistics is of Poisson type, which can be understood by noting that this system is close to the integrable limit where $V(x-y) = \delta(x-y)$. Nevertheless, we expect that even when $V(x)$ is only slightly detuned from the integrable limit, sufficiently large Hilbert space sectors will show thermalization (for the parameter choice in Fig.~\ref{fig: lvlstat3}, these are not accessible in our numerics). For all cases, we have chosen interaction potentials that are tuned to give rise to frugal exact states, whose presence does not affect level statistics considerations. Moreover, we have set $a \sim L \langle V \rangle/P_\mathrm{tot}$ for the quadratic part of the dispersion relation, which here is given by $\epsilon(p) = vp + ap^2$. This choice is made so that the non-linearity $a$ competes with the interaction $V(p)$: for $a \gg L \langle V \rangle/P_\mathrm{tot}$, the system approaches the free fermion integrable limit; for $a \ll L \langle V \rangle/P_\mathrm{tot}$, the system approaches the free boson limit. In either case, the level statistics approaches Poisson statistics unless $a \sim L \langle V \rangle/P_\mathrm{tot}$.

\section{Non-linearity in the bosonic basis} \label{app: non-linear-boson}
Here, we derive the Hamiltonian of the CNLLL in the bosonic basis (see also Sec.~\ref{sec: exact_boson}). We focus on the quadratic perturbation to the dispersion relation and neglect all higher-order terms in Eq.~\eqref{eq: vanillavanillaHamiltonian}. Restricting to a quadratic perturbation gives rise to a special algebraic structure (Eq.~\eqref{eq: gRSGA} below) that speeds up the numerical generation of Hamiltonian matrix elements for large Hilbert space sectors. For the sake of generality, in this appendix section only, we do not set $L=2\pi$, but instead keep $2\pi/L\equiv\Delta$ as a variable. We then consider the Hamiltonian
\begin{equation} \label{eq: quadHamiltonianFull}
\begin{aligned}
H &= \sum_{p} (vp+ap^2) c^\dagger_p c_p + H_\mathrm{int}
\\&\equiv H_\mathrm{lin} + H_\mathrm{quad} + H_\mathrm{int}, \quad H_\mathrm{lin} = v \sum_p p c^\dagger_p c_p, \quad H_\mathrm{quad} = a \sum_p p^2 c^\dagger_p c_p,
\end{aligned}
\end{equation}
where we assume $a p^2 \ll vp$. We begin by calculating
\begin{equation}
[H_\mathrm{quad}, b^\dagger_q] = a q^2 b^\dagger_q + 2 a \sqrt{\Delta q} \sum_{p} p c^\dagger_{p+q} c_p,
\end{equation}
implying that the spectrum-generating algebra of Eq.~\eqref{eq: HkinSGA} does not generalize to $H_\mathrm{quad}$. However, we furthermore have that
\begin{equation}
[[H_\mathrm{quad}, b^\dagger_q], b^\dagger_k] = 2a \sqrt{\Delta} \sqrt{k^2 q + k q^2} b^\dagger_{k+q},
\end{equation}
and therefore the algebra closes:
\begin{equation}
[[[H_\mathrm{quad}, b^\dagger_q], b^\dagger_k], b^\dagger_{l}] = 0.
\end{equation}
For the complete Hamiltonian $H$, we obtain
\begin{align}
&\relax[H, b^\dagger_q] = q \left[v + a q + V(q)\right] b^\dagger_q + 2 a \sqrt{\Delta q} \sum_{p} p c^\dagger_{p+q} c_p, \label{eq: gRSGA} \\
&[[H, b^\dagger_q], b^\dagger_k] = 2 a\sqrt{\Delta} \sqrt{k^2 q + k q^2} b^\dagger_{k+q}, \\
&[[[H, b^\dagger_q], b^\dagger_k], b^\dagger_{l}] = 0.
\end{align}
Now, consider the non-interacting ground state $\ket{\Omega}$, which is also still an eigenstate of the full Hamiltonian in that it satisfies
\begin{equation}
\normord{H} \ket{\Omega} = 0.
\end{equation}
Because $[H_\mathrm{quad}, b^\dagger_q]$ is not proportional to $b^\dagger_q$, we cannot anymore write down the full spectrum as we did in Eq.~\eqref{eq: fullbosonic_solution}. This breakdown of integrability is also implied by the commutator
\begin{equation}
[H_\mathrm{quad}, H_\mathrm{int}] \neq 0
\end{equation}
being non-zero, so that we cannot find a common diagonal basis for the kinetic and interacting parts of the Hamiltonian. 

To obtain the Hamiltonian in all Hilbert space sectors, labelled by total momentum \emph{and} the total particle number in Eq.~\eqref{eq: totalparticlenumop}, we now introduce $\ket{\Omega_N}$ as the ground state of $H_\mathrm{lin}$ at particle number $\normord{\hat{N}} \ket{\Omega_N} = N \ket{\Omega_N}$, so that (recall $\Delta=2\pi/L$)
\begin{equation}
\begin{aligned}
c^\dagger_p \ket{\Omega_N} = 0 \quad (p \leq \Delta N), \\
c_p \ket{\Omega_N} = 0 \quad (p > \Delta N).
\end{aligned}
\end{equation}
With respect to $\ket{\Omega}$, it has an energy
\begin{equation}
\normord{H_\mathrm{lin}} \ket{\Omega_N} = \frac{\Delta v}{2} N (N+1) \ket{\Omega_N},
\end{equation}
which is zero for $N=-1$ and otherwise positive.
Then, all other states at particle number $N$ are spanned by the orthonormal basis
\begin{equation}
\ket{N,\bs{m}} \equiv \prod_{q>0} \frac{b^{\dagger m_q}_q}{\sqrt{m_q!}} \ket{\Omega_N},
\end{equation}
where $0 \leq m_{q} \in \mathbb{Z}$, $q > 0$, are boson occupation numbers (we have that $b^\dagger_q b_q \ket{\Omega_N} = 0$ for all $N$).

We need to systematically express also the second term of $[H, b^\dagger_q]$, Eq.~\eqref{eq: gRSGA}, in terms of the bosonic and particle number operators. Such a representation must be possible because that term preserves $\hat{N}$, and the bosonic operators span the full Hilbert space at fixed $\normord{\hat{N}} = N$, see for example Eq.~\eqref{eq: bosonfermiunitaryexample}. In fact, we only need to find a representation of 
\begin{equation}
\sum_{p} p c^\dagger_{p+q} c_p \ket{\Omega_N} \equiv \sum_{\sum_l l m_l = q} \gamma_{q N \bs{m}} \ket{N,\bs{m}} = \sum_{\sum_l l m_l = q} \gamma_{qN\bs{m}} \prod_{l>0} \frac{b^{\dagger m_l}_l}{\sqrt{m_l!}} \ket{\Omega_N},
\end{equation}
where $q = \Delta \hat{q}$, and $\sum_l l m_l = \hat{q}$ is a partition of $\hat{q}$ into positive integers $\hat{l} m_l$. The contributions of all other boson occupations $\bs{m}$ vanish because the total momentum is fixed to $q$ -- we therefore treat $\bs{m}$ as an $\hat{q}$-dimensional vector, because its non-zero components may range from $m_1$ to $m_{\hat{q}}$. The expansion coefficients $\gamma_{q N \bs{m}}$ are obtained as
\begin{equation}
\gamma_{q N \bs{m}} = \bra{\Omega_N}\prod_{l>0} \frac{b^{m_l}_l}{\sqrt{m_l!}} \sum_{p} p c^\dagger_{p+q} c_p \ket{\Omega_N}.
\end{equation}
To evaluate them, we need the commutator
\begin{equation}
\begin{aligned}
\left[b_l, \sum_p p c^\dagger_{p+q} c_p\right] &= \sqrt{\frac{2\pi}{L l}} \sum_p p \left(c^\dagger_{p+q-l} c_p - c^\dagger_{p+q} c_{p+l}\right)
\\&= \sqrt{\frac{2\pi}{L l}} \sum_p \left( p \normord{c^\dagger_{p+q-l} c_p} + p \braket{\Omega_N |c^\dagger_{p+q-l} c_p|\Omega_N} - p \normord{c^\dagger_{p+q} c_{p+l}} - p \braket{\Omega_N | c^\dagger_{p+q} c_{p+l} |\Omega_N} \right)
\\&= \sqrt{\frac{2\pi}{L l}} \sum_p \left[ p \normord{c^\dagger_{p+q-l} c_{p}} - (p-l) \normord{c^\dagger_{p+q-l} c_{p}} + p \delta_{lq} \left(\delta_{\hat{p}\leq N}-\delta_{\hat{p}\leq -\hat{q}+N}\right) \right]
\\&= \sqrt{\frac{2\pi l}{L}} \sum_p \normord{c^\dagger_{p+q-l} c_{p}} + \frac{\delta_{lq} \Delta}{\sqrt{\hat{q}}} \sum_{-\hat{q}+N < \hat{p} \leq N} \hat{p}
\\&= \sqrt{l (q-l)} b^\dagger_{q-l} + \delta_{l q} \Delta \sqrt{\hat{q}} \left[N - \frac{1}{2}\left(\hat{q} -1 \right) \right],
\end{aligned}
\end{equation}
where we have taken care to only shift summation variables in normal-ordered expressions, and have used that $\braket{\Omega_N | c^\dagger_p c_q|\Omega_N}= \delta_{pq} \delta_{\hat{p} \leq N}$. Here, $\delta_{\hat{p} \leq N}$ is equal to $1$ when $\hat{p} \leq N$ and zero otherwise. Because of $[b_l, b^\dagger_p] = \delta_{lp}$ and $b_l \ket{\Omega_N} = 0$ for all $N$, only three distinct kinds of partitions contribute:

(1)~The trivial partition $\bs{m} = (0\dots, 1)$, with $m_k = \delta_{k q}$, comes with the coefficient
\begin{equation}
\begin{aligned}
\gamma_{q N \delta_{k q}} =& \bra{\Omega_N} b_q \sum_{p} p c^\dagger_{p+q} c_p \ket{\Omega_N} = \bra{\Omega_N} \left[b_q,\sum_{p} p c^\dagger_{p+q} c_p \right]\ket{\Omega_N} \\
=& \Delta \sqrt{\hat{q}} \left[N - \frac{1}{2}\left(\hat{q} -1 \right) \right].
\end{aligned}
\end{equation}

(2)~The bipartitions $\bs{m} = (0\dots, 1, 0\dots,1,0\dots)$, with $m_k = \delta_{k l} + \delta_{k,q-l}$, $l < q$, $l \neq q/2$, come with the coefficients
\begin{equation}
\begin{aligned}
\gamma_{q N, \delta_{k l} + \delta_{k,q-l}} =& \bra{\Omega_N} b_{q-l} b_l \sum_{p} p c^\dagger_{p+q} c_p \ket{\Omega_N} = \bra{\Omega_N} b_{q-l} \left[b_l, \sum_{p} p c^\dagger_{p+q} c_p \right]\ket{\Omega_N}
\\=& \bra{\Omega_N} b_{q-l} \sqrt{l (q-l)} b^\dagger_{q-l} \ket{\Omega_N} = \sqrt{l (q-l)}.
\end{aligned}
\end{equation}

(3) In the case where $\hat{q}$ is even, there is a symmetric bipartition that needs to be treated separately: $\bs{m} = (0\dots, 2, 0\dots)$, with $m_k = 2 \delta_{2k,q}$, has coefficient
\begin{equation}
\begin{aligned}
\gamma_{q N, 2 \delta_{2k,q}} =& \bra{\Omega_N} \frac{b^2_{q/2}}{\sqrt{2!}} \sum_{p} p c^\dagger_{p+q} c_p \ket{\Omega_N} = \sqrt{\frac{(q/2) (q/2)}{2}} = \frac{q}{\sqrt{8}}.
\end{aligned}
\end{equation}
In summary, we have derived that
\begin{equation} \label{eq: expressingFermionToBosonInCommutator}
\sum_{p} p c^\dagger_{p+q} c_p \ket{\Omega_N} = \Delta \sqrt{\hat{q}} \left[N - \frac{1}{2}\left(\hat{q} -1 \right) \right] b^\dagger_q \ket{\Omega_N} + \sum_{1 \leq \hat{l} < \hat{q}/2} \sqrt{l (q-l)} b^\dagger_{l} b^\dagger_{q-l} \ket{\Omega_N} + \frac{q}{\sqrt{8}} \frac{b^{\dagger 2}_{q/2}}{\sqrt{2}!} \ket{\Omega_N},
\end{equation}
where the last term is only to be included when $\hat{q}$ is even.
We also need 
\begin{equation} \label{eq: HactingonOMEGA}
\normord{H} \ket{\Omega_N} = \left(\normord{H_\mathrm{lin}} + \normord{H_\mathrm{quad}}\right) \ket{\Omega_N} = \frac{\Delta}{2} \left[v + \frac{\Delta a}{3} (2N+1)\right] N (N+1) \ket{\Omega_N} \equiv E_{0N}\ket{\Omega_N},
\end{equation}
so that the $N$-particle ground states $\ket{\Omega_N}$ already provide us with an exact tower of eigenstates at energies $E_{0N}$. We are now also in a position to write down the full Hamiltonian in the bosonic basis. We begin with
\begin{equation} \label{eq: HactingonbDaggerq}
\begin{aligned}
\normord{H} b^\dagger_q \ket{\Omega_N} &= \left([\normord{H}, b^\dagger_q] + E_{0N} b^\dagger_q \right)\ket{\Omega_N} = (E_{0N}+E_{\hat{q}}) b^\dagger_q \ket{\Omega_N} + 2 a \sqrt{\Delta q} \sum_{p} p c^\dagger_{p+q} c_p \ket{\Omega_N} \\
&\equiv E_{Nq} b^\dagger_q \ket{\Omega_N} + 2 a \sqrt{\Delta q} \left\{\Delta \sqrt{\hat{q}} \left[N - \frac{1}{2}\left(\hat{q} -1 \right) \right] b^\dagger_q + \sum_{1 \leq \hat{l} < \hat{q}/2} \sqrt{l (q-l)} b^\dagger_{l} b^\dagger_{q-l} + \frac{q}{\sqrt{8}} \frac{b^{\dagger 2}_{q/2}}{\sqrt{2}!}\right\}  \ket{\Omega_N}
\end{aligned}
\end{equation}
For closure, we need to complement this with the set of equations
\begin{equation}
\normord{H} b^\dagger_l b^\dagger_{q-l} \ket{\Omega_N} = \left\{[[\normord{H},b^\dagger_l],b^\dagger_{q-l}] + b^\dagger_{q-l} [\normord{H},b^\dagger_l] +b^\dagger_l [\normord{H}, b^\dagger_{q-l}] + E_{0N} b^\dagger_l b^\dagger_{q-l} \right\} \ket{\Omega_N}.
\end{equation}
Plugging in the commutators from Eq.~\eqref{eq: gRSGA}, using Eq.~\eqref{eq: expressingFermionToBosonInCommutator}, and iterating, we find that $\normord{H}$ explores the space of all partitions, because we can write every partition in terms of successive bipartitions.

Now, first note that Eqs.~\eqref{eq: HactingonOMEGA} and~\eqref{eq: HactingonbDaggerq} can be rewritten as 
\begin{equation}
\begin{aligned}
\normord{H} \ket{\Omega_N} &= E(N,0) \ket{\Omega_N}, \\
\normord{H} b^\dagger_q \ket{\Omega_N} &= \left[E(N,q) b^\dagger_q + \sum_{l < q/2} G(l,q) b^\dagger_l b^\dagger_{q-l} + B(q) \frac{b^{\dagger 2}_{q/2}}{\sqrt{2}!} \right] \ket{\Omega_N},
\end{aligned}
\end{equation}
where we have introduced the functions
\begin{equation}
\begin{aligned}
E(N,q) &= v \Delta \frac{N(N+1)}{2} + a \Delta^2 \frac{N(N+1)(2N+1)}{6} + q \left[v+V(q) + a \Delta (2N+1)\right],
\\
G(l,q) &= 2 a \sqrt{l q(q-l)\Delta}, \\ B(q) &= 2 a \sqrt{\frac{q^3 \Delta}{8}}.
\end{aligned}
\end{equation}
Furthermore, all higher-order terms can be obtained by commutator iteration (recursion relation):
\begin{equation}
\begin{aligned}
\normord{H} b^\dagger_{q_{n+2}} b^\dagger_{q_{n+1}} b^\dagger_{q_n} \dots b^\dagger_{\Delta} \ket{\Omega_N} = 
&2 a \sqrt{q_{n+2}q_{n+1}(q_{n+2}+q_{n+1})\Delta} b^\dagger_{q_{n+2} + q_{n+1}} b^\dagger_{q_n} \dots b^\dagger_{\Delta} \ket{\Omega_N} \\
&+(b^\dagger_{q_{n+2}} H b^\dagger_{q_{n+1}} + b^\dagger_{q_{n+1}} H b^\dagger_{q_{n+2}}) b^\dagger_{q_n} \dots b^\dagger_{\Delta} \ket{\Omega_N} \\
&-b^\dagger_{q_{n+2}} b^\dagger_{q_{n+1}} H b^\dagger_{q_n} \dots b^\dagger_{\Delta} \ket{\Omega_N}
\end{aligned}
\end{equation}
The Hamiltonian matrix is now fully specified and can be generated using a computer algebra system.

\section{Example of an infinite family of non-frugal sequences} \label{sec: generalizable_sequences_appendix}
In generalization to the analysis in Sec.~\ref{sec: generalizable_sequences}, we here write down an infinite family of exact states that are non-frugal in general, given by 
\begin{equation} \label{eq: infinitesequence1}
\begin{aligned}
&\ket{\Psi^{(+2m,1)}_{\bar{k}}} = c^\dagger_{2\bar{k}+2m} \left(\prod_{k'=1}^{\bar{k}-1} c^\dagger_{2k'} \right) \left(\prod_{k=1}^{\bar{k}} c_{-(2k-1)} \right) \ket{\Omega}, \\
&\ket{\Psi^{(+2m,1)}_{\tilde{k}}} = c^\dagger_{2\tilde{k}-1+2m} \left(\prod_{k'=0}^{\tilde{k}-2} c^\dagger_{2k'+1}\right) \left(\prod_{k=0}^{\tilde{k}-1} c_{-2k} \right) \ket{\Omega}.
\end{aligned}
\end{equation}
These states are the generalization of Eq.~\eqref{eq: firstdescendants}, where we increase the momentum of the highest occupied state by $2m$ rather than $2$. Similarly, we can build the analogue of the eigenstates in Eq.~\eqref{eq: seconddescendants} by reducing the momentum of the lowest unoccupied state by $2m$ rather than $2$:
\begin{equation} \label{eq: infinitesequence2}
\begin{aligned}
\ket{\Psi^{(+2m,2)}_{\bar{k}}} =& \left(\prod_{k'=1}^{\bar{k}} c^\dagger_{2k'} \right) \left(\prod_{k=1}^{\bar{k}-1} c_{-(2k-1)} \right) c_{-(2\bar{k}-1)-2m} \ket{\Omega}, \\
\ket{\Psi^{(+2m,2)}_{\tilde{k}}} =& \left(\prod_{k'=0}^{\tilde{k}-1} c^\dagger_{2k'+1} \right) \left(\prod_{k=0}^{\tilde{k}-2} c_{-2k} \right) c_{-2(\tilde{k}-1)-2m} \ket{\Omega}.
\end{aligned}
\end{equation}
These states are defined for any $m \in \mathbb{N}$, where $m=0$ corresponds to Eqs.~\eqref{eq: statesexactsequence1} and~\eqref{eq: statesexactsequence2}, while $m=1$ corresponds to Eqs.~\eqref{eq: firstdescendants} and~\eqref{eq: seconddescendants}. They have total momentum $P=2m+\bar{k} (2\bar{k}+1)$ and $P=2m+\tilde{k}(2\tilde{k}-1)$, respectively, and are exact eigenstates of $\normord{H}$ in Eq.~\eqref{eq: Hint_vanilla} as long as the potentials $V(2) = V(4) = \dots = V(2m) = V(1) = V(3) = V(5) = \dots = V(2(n+m)+1)$ are set equal. We note that this is a self-consistent potential configuration (Sec.~\ref{sec: selfconsistency}), and indeed the states in Eqs.~\eqref{eq: infinitesequence1} and~\eqref{eq: infinitesequence2} are of the form of Eq.~\eqref{eq: selfconsistent_eigenstate} and its particle-hole dual. From the data presented in App.~\ref{app: exact_tally}, we find that there are no frugal examples for $m > 1$ in the total momentum sectors with $P_{\mathrm{tot}} \leq 40$.

\section{Exact state entanglement in real space} \label{app: real_space_entanglement}

In this appendix, we explain how to calculate the entanglement entropy of momentum-space Slater-determinant states in \emph{continuous} real space by diagonalizing a finite-dimensional matrix. This method was used to plot Fig.~\ref{fig: entanglement}c and is based on Refs.~\onlinecite{Rezayi12,Bonderson12} and on the formula for the entanglement entropy for Gaussian states~\cite{Peschel_2003}.

\subsection{Single-particle states}
We assume periodic boundary conditions on a one-dimensional ring of circumference $L$ (in the main text, we have set $L = 2\pi$). The momentum-space orbitals are given by 
\begin{equation}
\ket{\Psi_p} = \frac{1}{\sqrt{L}} \int_{-L/2}^{L/2} \mathrm{d}x \, e^{\mathrm{i} p x} \ket{x},
\end{equation}
where $p \in 2\pi \mathbb{Z}/L$, and $\ket{x} = c^\dagger_x \ket{0}$ is a basis state at position $x$ where $\braket{x|y} = \delta(x-y)$ (see Eqs.~\eqref{eq: ccdagdef} and~\eqref{eq: CAR}), while $\ket{0}$ is the fermionic vacuum. Then, the momentum-space orbitals satisfy $\braket{\Psi_p | \Psi_q} = \delta_{pq}$. 
For a real-space decomposition $[-L/2,L/2] = A \cup B$ without overlap between $A$ and $B$, we introduce $\ket{a}$ and $\ket{b}$ as orthonormal orbitals with exclusive support in subregions $A$ and $B$, respectively:
\begin{equation} \label{eq: conditions_on_orbitals}
\mathcal{P}_A \ket{a} = \ket{a}, \quad \mathcal{P}_B \ket{a} = 0, \quad \braket{a|a'} = \delta_{a a'}, \quad
\mathcal{P}_B \ket{b} = \ket{b}, \quad \mathcal{P}_A \ket{b} = 0, \quad \braket{b|b'} = \delta_{b b'}, \quad \braket{a|b}=0,
\end{equation}
where 
\begin{equation}
\mathcal{P}_{A,B} = \int_{A,B} \mathrm{d}x \, \ket{x} \bra{x}
\end{equation}
is the projector onto subregion $A$ or $B$. Now, we may decompose each momentum-space orbital as
\begin{equation}
\ket{\Psi_p} = \sum_{a} \mathcal{T}^{A}_{p a} \ket{a} + \sum_{b} \mathcal{T}^{B}_{p b} \ket{b}, \quad \mathcal{T}^{A}_{p a} = \braket{a | \Psi_p}, \quad \mathcal{T}^{B}_{p b} = \braket{b | \Psi_p}.
\end{equation}

To fully capture $|p|$ plane waves $\ket{\Psi_p}$, we must at most consider a total of $|a|, |b| \leq |p|$ orbitals in either subregion. Since in continuous space, no finite number $|p|$ of plane waves spans the full space of single-particle states, there is \emph{no} completeness constraint, \emph{i.e.}, $|a| + |b|$ is not necessarily equal to $|p|$, so that the $\mathcal{T}$ matrix only acts unitarily in the Hilbert subspace spanned by the $|p|$ plane waves. 
We now introduce the Hermitian $|p| \times |p|$ matrix of plane-wave overlaps restricted to subregion $A$,
\begin{equation} \label{eq: specdecompOmat}
(\mathcal{O}_A)_{p q} = \braket{\Psi_p |\mathcal{P}_A |\Psi_q} \equiv \sum_{i} \lambda_i v_{i p} \bar{v}_{i q},
\end{equation}
where in the last equation we introduced the spectral decomposition of $\mathcal{O}_A$, using its eigenvalues $\lambda_i \geq 0$, $i = 1 \dots |p|$, and eigenvectors $v_{i p}$. For the subregion $A = [-L_A/2,L_A/2]$, we find
\begin{equation} \label{eq: specificOmatrix}
(\mathcal{O}_A)_{p q} = \frac{1}{L} \int_{-L_A/2}^{L_A/2} \mathrm{d}x \int_{-L_A/2}^{L_A/2} \mathrm{d}y \, e^{-\mathrm{i} (p x - q y)} \braket{x | y} = \frac{2}{L(p-q)} \sin\left[\frac{L_A (p-q)}{2}\right],
\end{equation}
where $p$ and $q$ only range over occupied momenta. In practice, we diagonalize $\mathcal{O}_A$ numerically to find $\lambda_i$ and $v_{i p}$.

Making use of the spectral decomposition of $\mathcal{O}_A$ in Eq.~\eqref{eq: specdecompOmat}, a convenient choice of orbitals for subregion $A$ satisfying all requirements of Eq.~\eqref{eq: conditions_on_orbitals} is then given by
\begin{equation} \label{eq: atrialstate}
\ket{a}=\sum_p \frac{v_{a p}}{\sqrt{\lambda_{a}}} \mathcal{P}_A \ket{\Psi_p}, \quad \mathcal{P}_A \ket{\Psi_p} = \sum_{a} \sqrt{\lambda_a} \bar{v}_{a p} \ket{a},
\quad \rightarrow \quad \mathcal{T}^{A}_{p a} = \sqrt{\lambda_a} \bar{v}_{a p},
\end{equation}
where $a$ only ranges over the non-zero eigenvalues $\lambda_a > 0$ of $\mathcal{O}_A$. Therefore, if $\mathcal{O}_A$ has zero-modes because the restricted orbitals $\mathcal{P}_A \ket{\Psi_p}$ are not linearly independent in subregion $A$, then $|a|$ is strictly smaller than $|p|$. To derive the second equation in~\eqref{eq: atrialstate}, we have used that $\sum_p v_{j p} \mathcal{P}_A \ket{\Psi_p} = 0$ holds when $\lambda_j = 0$, that is, for all zero-modes of $\mathcal{O}_A$.
Now, because
\begin{equation}
\sum_{i} \left(\sqrt{\lambda_i} \bar{v}_{p i}\right)^\dagger \left(\sqrt{\lambda_i} \bar{v}_{i q}\right) = (\mathcal{O}_A)_{p q},
\end{equation}
we can identify 
\begin{equation}
\left(\sqrt{\lambda_i} \bar{v}_{i q}\right) = \left(U \mathcal{O}_A^{1/2}\right)_{i q},
\end{equation}
where $U$ is a unitary matrix and $\mathcal{O}_A^{1/2}$ denotes the unique Hermitian square root of the Hermitian positive-semidefinite matrix $\mathcal{O}_A$. We will see that $U$ does not enter in the calculation of the entanglement entropy. In conclusion, we have derived
\begin{equation}
\mathcal{P}_A \ket{\Psi_p} = \sum_{a} \left(U \mathcal{O}_A^{1/2}\right)_{a p} \ket{a},
\end{equation}
where $\ket{a}$ forms an orthonormal basis for subregion $A$. A similar construction applies to the orbitals in subregion $B$.

\subsection{Slater-determinant state entanglement}
For a given set of occupied momenta $p \in \mathrm{occ}$, the restricted correlation matrix, which is used to calculate the entanglement entropy of Slater-determinant states~\cite{Peschel_2003}, is given by 
\begin{equation} \label{eq: restrictedcorrmat}
\mathcal{P}_A \mathcal{P}_\mathrm{occ} \mathcal{P}_A = \sum_{p \in \mathrm{occ}} \mathcal{P}_A \ket{\Psi_p} \bra{\Psi_p} \mathcal{P}_A = \sum_{a a'} \tilde{\mathcal{C}}^{\mathrm{occ}}_{a a'} \ket{a}\bra{a'},
\end{equation}
where we have defined
\begin{equation}
\tilde{\mathcal{C}}^{\mathrm{occ}}_{a a'} = \sum_{p \in \mathrm{occ}} \left(U \mathcal{O}_A^{1/2}\right)_{a p} \left(\mathcal{O}_A^{1/2} U^\dagger \right)_{p a'}.
\end{equation}
Now, because $\ket{a}$ is an orthonormal basis in subregion $A$, we can diagonalize the $|a| \times |a|$ matrix $\tilde{\mathcal{C}}^{\mathrm{occ}}$, $|a| \leq |\mathrm{occ}|$, to find all non-zero entanglement eigenvalues. Specifically, let $\lambda_\alpha$, $\alpha = 1 \dots |a|$, be the eigenvalues of $\tilde{\mathcal{C}}^{\mathrm{occ}}$. Since $U$ is unitary, they are equal to the eigenvalues of the matrix
\begin{equation} \label{eq: cmatrix}
\mathcal{C}^{\mathrm{occ}}_{a a'} = \sum_{p \in \mathrm{occ}} \left(\mathcal{O}_A^{1/2}\right)_{a p} \left(\mathcal{O}_A^{1/2} \right)_{p a'},
\end{equation}
so that $U$ drops out and only knowledge of $\mathcal{O}_A^{1/2}$ is required. Then, the entanglement entropy is given by~\cite{Peschel_2003}
\begin{equation}
S_A = - \sum_{\alpha} \left[ \lambda_\alpha \log \lambda_\alpha + (1-\lambda_\alpha) \log (1-\lambda_\alpha) \right].
\end{equation}
This implies that $S_A$ can be calculated using the $\mathcal{O}_A$ matrix itself, even if it has zero-modes that in principle reduce the number of orbitals $\ket{a}$ needed in subregion $A$. 

Note that we are dealing with a chiral one-dimensional system, thus precluding a lattice representation. Moreover, working directly with $\mathcal{P}_A \mathcal{P}_\mathrm{occ} \mathcal{P}_A$ in real space would in principle require diagonalizing an infinite-dimensional matrix. This difficulty is circumvented by the introduction of an orthonormal basis in subregion $A$ encoding the chiral momentum constraint.

\subsection{Symmetries of the entanglement entropy}
Remarkably, the entanglement entropy of the exact eigenstates discussed in the main text is invariant under the particle-hole duality defined in Sec.~\ref{sec: duality}, even when their energy is not. In Eq.~\eqref{eq: duality}, the duality transformation is obtained by first reflecting all momenta $p \rightarrow 1-p$, and then implementing a particle-hole conjugation $n_p \rightarrow \bar{n}_p$. We now recall that the entanglement entropy of a Slater-determinant state is calculated from the eigenvalues of its restricted correlation matrix~\cite{Peschel_2003}, Eq.~\eqref{eq: restrictedcorrmat},
\begin{equation}
C = \mathcal{P}_A \mathcal{P}_\mathrm{occ} \mathcal{P}_A.
\end{equation}
The reflection $p \rightarrow 1-p$ is composed of two parts: (1) a spinless time-reversal symmetry sending $p \rightarrow -p$, and represented in Hilbert space by an anti-unitary operator $T$, and (2) a momentum shift sending $p \rightarrow p + 1$, which can be absorbed by a re-definition of the momentum coordinate origin as long as the momentum-space cutoff $\Lambda$ (defined in Sec.~\ref{subsec: hamiltonian_first}) is sent to infinity. Using that time-reversal symmetry leaves the real-space projector $\mathcal{P}_A$ invariant, we have
\begin{equation}
\mathcal{P}_A \left(T \mathcal{P}_\mathrm{occ} T^\dagger\right) \mathcal{P}_A = T C T^\dagger,
\end{equation}
which has the same spectrum as $C$ because $T$ is anti-unitary ($C$ is Hermitian and hence has a real spectrum). Thus, the reflection $p \rightarrow 1-p$ leaves the entanglement entropy unaffected as long as the momentum-space cutoff $\Lambda$ is sent to infinity.
Next, it was shown in Ref.~\onlinecite{Lai15} that the spectrum of $C$ is identical to the spectrum of the matrix
\begin{equation}
\tilde{C}_{\mathrm{occ}} = \mathcal{P}_\mathrm{occ} \mathcal{P}_A \mathcal{P}_\mathrm{occ}.
\end{equation}
This momentum-space--real-space duality implies that the following two entanglement entropies are identical:
\begin{itemize}
\item[(1)]{the entanglement entropy of a Slater-determinant state with projection operator $\mathcal{P}_\mathrm{occ}$, in presence of an entanglement cut separating subregion $A$ from its complement in real space, and}
\item[(2)]{the entanglement entropy of a Slater-determinant state with projection operator $\mathcal{P}_A$, in presence of an entanglement cut separating the occupied subspace of $\mathcal{P}_\mathrm{occ}$ from its complement in momentum space.}
\end{itemize}
The particle-hole conjugation $n_p \rightarrow \bar{n}_p$ implies a substitution $\mathcal{P}_\mathrm{occ} \rightarrow \mathcal{P}_\mathrm{emp}$, where $\mathcal{P}_\mathrm{emp} = \mathbb{1} - \mathcal{P}_\mathrm{occ}$ is the projector onto the empty subspace. Hence, the particle-hole conjugated entanglement entropy may be obtained from
\begin{equation}
\tilde{C}_{\mathrm{emp}} = \mathcal{P}_\mathrm{emp} \mathcal{P}_A \mathcal{P}_\mathrm{emp}.
\end{equation}
Since $\mathcal{P}_A$ projects onto the occupied subspace of a pure quantum state -- the Slater-determinant state that has all real-space orbitals in subregion $A$ occupied -- the entanglement entropy is actually independent of whether the empty or occupied subspace is traced out. Hence, the entanglement entropy calculated from $\tilde{C}_{\mathrm{emp}}$ is identical to the entanglement entropy calculated from $\tilde{C}_{\mathrm{occ}}$.

In conclusion, we find that the transformation in Eq.~\eqref{eq: duality} does not change the entanglement entropy of a chiral Slater-determinant state, as long as the momentum-space cutoff $\Lambda$ is sent to infinity. We note, however, that numerical calculations necessitate a finite cutoff, and so a small mismatch remains at any finite cutoff. This mismatch is invisible to the eye in Fig.~\ref{fig: entanglement}, which uses $\Lambda = 100$.

\section{Naive estimate for the maximal real-space entanglement entropy in chiral systems} \label{sec: naive_SA_estimate}
To estimate the maximal real-space entanglement entropy of excited states, we first note that the states of a chiral system in principle involve an infinite number of occupied orbitals at negative momenta. In practice, we introduce a cutoff $-\Lambda \ll - P_\mathrm{tot} \leq 0$, $\Lambda > 0$, and occupy by default all momenta at $-\Lambda < p \leq - P_\mathrm{tot}$, so that each fermionic state of the Hilbert space sector at $P_\mathrm{tot}$, given in Eq.~\eqref{eq: Slaterdef}, contains a total of $\Lambda$ electrons. Assuming a homogeneous charge distribution and a subregion length $L_A$, $A$ will host $N_A = \Lambda l_A$ electrons on average, where $l_A = L_A/L$. At the same time, the fermionic basis states of a given $P_\mathrm{tot}$ sector may access all momentum orbitals in the range $(-\Lambda,P_{\mathrm{tot}}]$. Since real space is continuous, any subregion $A$ -- independent of its size -- can in principle accommodate all $(P_\mathrm{tot}+\Lambda)$ orbitals without linear dependencies. Taking into account that a real-space cut preserves particle number, this observation yields an entanglement entropy bound
\begin{equation} \label{eq: naiveSAbound}
\begin{aligned}
S^{\mathrm{max}}_A &= \log \begin{pmatrix} P_\mathrm{tot}+\Lambda \\\Lambda l_A \end{pmatrix}
\\&\approx \Lambda \left[-l_A \log l_A - (1-l_A) \log (1-l_A)\right],
\end{aligned}
\end{equation}
which grows linearly with $\Lambda$. In the second line, we have assumed that $\Lambda \gg P_{\mathrm{tot}}$, as well as $\Lambda \gg 1$ and $\Lambda l_A \gg 1$. Note that $S^{\mathrm{max}}_A$ is closely related to the Page estimate~\cite{Page93,Rigol17,Bianchi19,Murthy19} of the average entanglement entropy. 

However, the entanglement entropy of an exact Slater-determinant eigenstate of the Hamiltonian in Eq.~\eqref{eq: Hint_vanilla} grows only logarithmically with $\Lambda$, in contrast to the linear growth of $S^{\mathrm{max}}_A$ in Eq.~\eqref{eq: naiveSAbound}: this is because any additional momentum orbitals at $-\Lambda \leq p \leq - P_\mathrm{tot}$ are completely filled by default, and hence -- employing the position-momentum duality of Ref.~\cite{Lai15} -- yield logarithmic contributions to the entanglement entropy at fixed $l_A$. 

We exemplify our previous discussion by considering $\Lambda = 100$ and the frugal state $\#12$ in the $P_\mathrm{tot}=20$ sector (Fig.~\ref{fig: entanglement}a and Fig.~\ref{fig: exactstates}a). We find $S_A \approx 4$ at $l_A=1/4$, while $S^{\mathrm{max}}_A \approx 59$. The discrepancy between $S^{\mathrm{max}}_A$ and the actual exact state entanglement can be understood by noting that our naive estimate does not properly take into account the effect of chirality. Indeed, combined with total momentum conservation, chirality leads to a finite Hilbert space at a given total momentum. However, any real space entanglement cut breaks translational symmetry -- hence, we lose total momentum as a good subregion quantum number, and cannot use it to further constrain Eq.~\eqref{eq: naiveSAbound}. Nevertheless, the constraints of chirality and total momentum conservation, not properly reflected in Eq.~\eqref{eq: naiveSAbound}, are expected to greatly reduce the average entanglement entropy. This is because, in chiral states, the vast majority of momenta (all momenta $p$ with $-\Lambda < p \leq - P_\mathrm{tot}$) is occupied homogeneously. To our knowledge, there is so far no analytic understanding of these effects on the typical behavior of the real-space entanglement entropy, which makes it difficult to assess how similar the exact states are to featureless thermal states. This situation should be put in contrast with the entanglement entropy of non-chiral free fermions, for which bounds based on lattice discretization (unavailable for chiral fermions) were established in Ref.~\onlinecite{Rigol17}. 

\end{document}